\documentclass[preprint,12pt]{elsarticle}

\usepackage{amssymb}
\usepackage{amsmath}
\usepackage{subcaption}
\usepackage{miller}
\usepackage{hyperref}
\usepackage{lineno}
\usepackage{subcaption}
\usepackage{graphicx}
\usepackage{xr}
\journal{Acta Materialia}

\begin{document}

\begin{frontmatter}



\title{Spall failure of alumina at high-strain rates using femtosecond laser experiments and high-fidelity molecular dynamics simulations}


\author[inst1]{Mewael Isiet}

\affiliation[inst1]{organization={Department of Mechanical Engineering},
            addressline={The University of British Columbia}, 
            city={Vancouver}, 
            state={British Columbia},
            postcode={V6T 1Z4},
            country={Canada}}
\author[inst1]{Musanna Galib}
\author[inst2,inst3]{Yunhuan Xiao}
\author[inst2,inst3]{Jerry I. Dadap}
\author[inst2,inst3]{Ziliang Ye}
\author[inst1]{Mauricio Ponga}

\affiliation[inst2]{organization={Department of Physics \& Astronomy},
            addressline={The University of British Columbia}, 
            city={Vancouver},
            state={British Columbia}, 
            postcode={V6T 1Z1},
            country={Canada}}

\affiliation[inst3]{organization={Quantum Matter Institute},
		addressline={The University of British Columbia}, 
		city={Vancouver},
		state={British Columbia}, 
        postcode={V6T 1Z1},		
		country={Canada}}
	
\begin{abstract}
Ceramic materials are widely used in high-strain-rate applications due to their exceptional strength-to-weight ratio. 
However, under these extreme conditions, spall failure becomes a critical concern, which is driven by a large hydrostatic tensile stress state. 
This study introduces a novel two-laser setup to generate controlled hydrostatic stress states at specific locations within test specimens. 
By inducing and manipulating shock wave interactions, we achieve large hydrostatic compressive and tensile stresses at very high-strain-rates, enabling the controlled nucleation and growth of nanovoids leading to spall failure. 
Our experiments demonstrate that shock wave interference can precisely trigger spallation at arbitrary locations in the specimen thickness. 
To further validate our approach, we investigate alumina spall failure using molecular dynamics (MD) simulations with a custom-designed graph neural network potential. 
The MD results show strong agreement with experimentally estimated spall strength. 
These findings highlight the potential of the two-laser technique as a powerful tool for studying the early stages of spall failure in ceramics, paving the way for advanced materials testing methodologies.
\end{abstract}

\begin{graphicalabstract}
	\includegraphics[width=\linewidth]{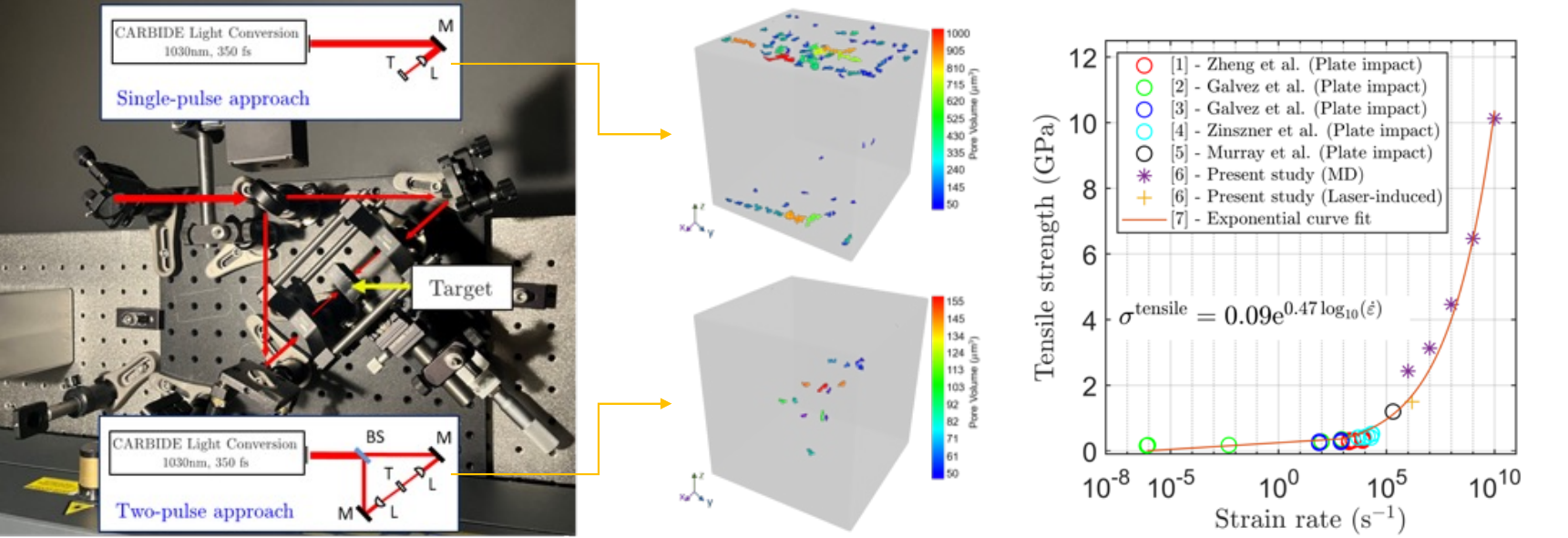}
\end{graphicalabstract}


\end{frontmatter}

\section{Introduction}
Ceramics have been extensively used in ballistic applications \cite{gooch2002overview, Medvedovski2010pt2,Silva2014,Liu2016} mainly due to their attractive strength to density ratio in comparison to metals \cite{Medvedovski2010pt1,Medvedovski2006}. 
The most commonly used ceramics are alumina ($\mathrm{Al_{2}O_{3}}$), silicon carbide (SiC), and boron carbide ($\mathrm{B_{4}C}$).
Amongst these ceramics, alumina due to its low cost and ability to be manufactured using a variety of methods, such as slip casting, powder compacting, injection molding, and additive techniques, has been the preferred material to be used in ballistic protection \cite{ApplebyThomas2020, Dresch2021}.
These advanced ceramics play an important role within an armor system against projectile impact by effectively dissipating ballistic energy.
Unlike metallic armor, which tends to absorb impact energy through plastic deformation via dislocation emission from nanovoids, ceramics tend to fracture to dissipate the kinetic energy of incoming projectiles \cite{Medvedovski2010pt1}.
Ceramics plates used in armor panels can typically experience high-velocity impacts ranging between 250 to 2000 $\mathrm{m} \cdot \mathrm{s}^{-1}$ \cite{Reaugh1999, Krell2014,Andraskar2022}, generating shock wave propagation at very high-strain-rates.
Upon impact, failure takes place leading to the perforation of the ceramic target.
These failures occur due to a combination of various mechanisms such as changes in the microstructural features, material strength, impact velocity and orientation, and projectile shape \cite{Backman1978}.
Ceramics, due to their brittle behavior, fail as a result of either plugging, spall failure, or radial fracture \cite{Yungwirth2011,Serjouei2015}.
Due to ceramics' lower tensile strength, compared to their compressive strength, failure due to tensile loading typically occurs first, i.e., spall failure \cite{Andraskar2022, Zinszner2015}.
Spall behavior of alumina has been extensively studied using the plate impact method \cite{Louro1989, Murray1998,Bourne2001,Girlitsky2014,Hayun2018,Lebar2021}.
Here, the impact of an accelerated flyer plate against a target is used to generate a large hydrostatic tensile state to initiate failure.
The overall consensus on the behavior of alumina from plate impact experiments is as follows: (i) spall resistance increases with increase in $\mathrm{Al_{2}O_{3}}$ content as impurities introduce flaws which could act as weak spots for the nucleation of voids, (ii) spall behavior is sensitive to pulse duration and initial microstructure, (iii) spall strength tends to decrease with increasing shock pressure due to incremental cracking of the material during the compression phase, and (iv) three failure modes have been observed under microscopy, i.e. fracture nucleation from voids, trans-granular micro-cracks, and inter-granular fracture.
Some degree of plastic activity (twinning and deformation bands) have also been identified \cite{Louro1989,Lankford1977}.
These results show the effectiveness of the conventional plate impact method; however, its destructive, time-intensive nature, need for large samples, and limited strain rates below $\mathrm{10^{6}~s^{-1}}$ make it unsuitable for high-throughput experiments and studying alumina's spall behavior at extremely high-strain-rates \cite{Mallick2021}.
Laser-driven shock methods, including the laser-induced impact test \cite{Thevamaran2016,Griesbach2023}, laser-driven micro-flyer \cite{Mallick2020,Wu2023}, and two-pulse laser shock \cite{Mewael_1st_paper}, have emerged as viable alternatives to conventional techniques.
These methods not only achieve strain rates exceeding $\mathrm{10^{6}~s^{-1}}$, but also offer advantages such as lower energy requirements for deforming smaller specimens and simplified procedures for aligning diagnostics.
However, since experimental spall studies are confined to free surface diagnostics and \textit{post-morten} observation \cite{Ponga2016, Grgoire2017}, it is necessary to incorporate computational methods to understand the failure mechanisms occurring at very small temporal (\textbf{fs}--\textbf{ns}) and spatial (\textbf{nm}) scales.
As a result, molecular dynamics (MD) has served as a competent tool to study the spall behavior of metals (such as Ni \cite{Shugaev2019,Ivanov2003}, Al \cite{Shao2013,Galitskiy2018}, and Cu \cite{Fensin2014,Xiong2017}), alloys (such as high-entropy alloys \cite{Thrmer2022,Li2023}, magnesium-based alloys \cite{Nitol2020,Yang2022}, and Ni-based alloys \cite{Wu2023,Chen2023}), and ceramics, such as $\mathrm{Al_{2}O_{3}}$ \cite{Chang_Hogan_2023,Zhang2008} and SiC \cite{Li2016,Li2017}.
Amongst the spall studies on ceramics, the Vashishtha \cite{Vashishta2011}, COMB3 \cite{Choudhary2015}, and ReaxFF \cite{Hong2016} potentials have been used to model material behavior, and are reactive, charge-and bond-order based.
While, generally speaking, these potentials allow the simulation of large and complex phenomena, their accuracy has been questioned for certain applications \cite{Wan2021, Piroozan2023} and as it is not straightforward to modify or extend such potentials readily \cite{Naserifar2013}, alternative approaches have been developed to remedy this. 
Allegro \cite{Batzner2022, Musaelian2023}, an equivariant graph neural network (GNN) based machine learning model, aims to approximate the potential energy surface using density functional theory (DFT) data to allow for the simulation of large-scale systems.
In this work, we focus on the application of the single- and two-pulse laser shock setup to study the spall behavior of commercially available alumina samples.
Typically in spall experiments, time-resolved diagnostics are performed using velocimetry to measure the free-surface velocity, and infer the spall strength \cite{Murray1998, Zinszner2015, Girlitsky2014}. 
However, in this work, the spall strength is estimated using the spall thickness captured via X-ray micro-computed tomography imaging.
Post-mortem analysis of shocked-alumina will be performed to understand the micro-structural failure mechanisms of ceramics.
\textit{In-situ} observation will be performed on molecular dynamics using a DFT-based machine learning potential to enable characterization of spallation in alumina ceramics.
\section{Methodology}
\subsection{Laser shock experimental setup}
\begin{figure}[ht]
	\centering
	\begin{subfigure}[b]{0.45\textwidth}
		\centering
		\includegraphics[width=\linewidth]{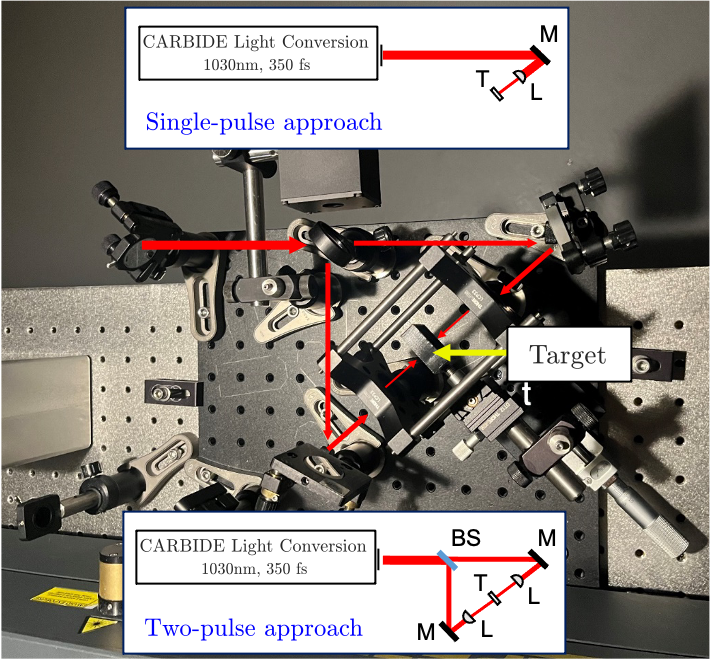}
		\caption{}
		\label{fig:experimental_layout}
	\end{subfigure}
	\hfill
	\begin{subfigure}[b]{0.45\textwidth}
		\centering
		\includegraphics[width=\linewidth]{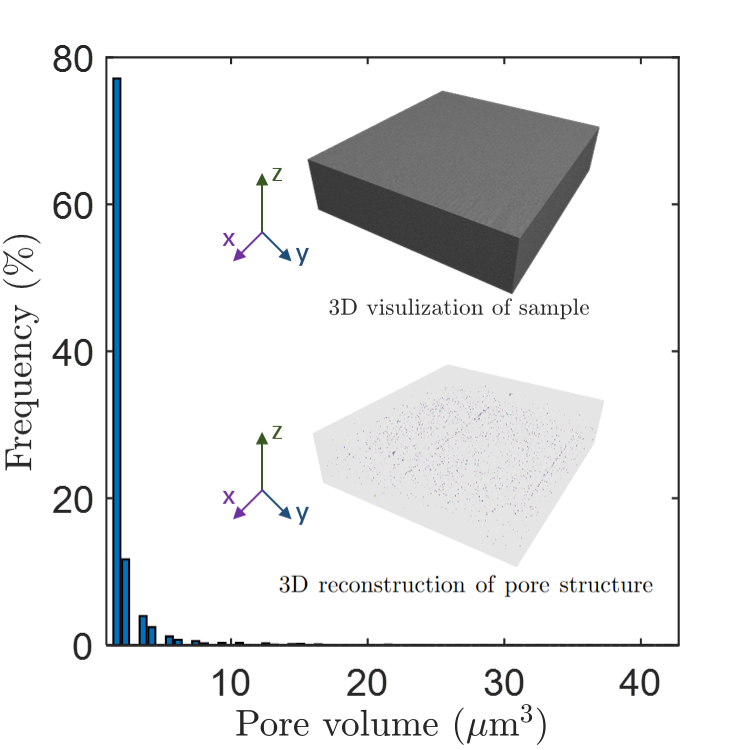}
		\caption{}
		\label{fig:pristine_histogram}
	\end{subfigure}
	\caption{Schematic of the (a) single- and two-pulse laser-induced spallation setups that illuminates either the front surface only or both the front and back surfaces and (b) the histogram of the pore volume of the pristine $\mathrm{Al_{2}O_{3}}$ sample with inset images of the micro-CT scan of the solid volume and pore structure.}
	\label{fig:experimental_layout_pristine_histogram}
\end{figure}

The experimental setup can be seen in Figure \ref{fig:experimental_layout} where the laser beam is directed to a single surface of the sample in the single-pulse approach or using a series of mirrors and a beam-splitter is used to illuminate the front and back surface of the sample in the two-pulse approach.
The ultra-short laser pulses at a maximum energy of 90 $\mu$J are generated by a femtosecond solid-state laser system (Carbide CB3-40W, Light Conversion) with a pulse duration of 350 fs.
Using the laser's internal shutter, a single laser pulse can be emitted with a full width at half maximum of 3.9 mm, and incorporating focusing lenses ensures that the spot size and the laser fluence can be controlled by careful positioning of the sample.
More details of the experimental setup can be found in our previous work \cite{Mewael_1st_paper}.
Commercially available $\mathrm{Al_{2}O_{3}}$ (Lithoz America) samples with 99.8$\%$ purity and a thickness of 250 $\mu$m were used in this study.
The porosity of the printed $\mathrm{Al_{2}O_{3}}$ was analyzed using an X-ray micro-computed tomography scan (micro-CT, Zeiss Xradia 520 Versa) with a voxel resolution of 1 $\mu m$. 
A 3D reconstruction of the sample revealed small voids, resulting in a porosity of less than 0.01$\%$ with a mean pore volume of 1.95 $\mu m^{3}$,  with the smallest, largest, and 95th percentile pore volumes measuring 1 $\mu m^{3}$, 42 $\mu m^{3}$, and 3.58 $\mu m^{3}$, as shown in Figure \ref{fig:pristine_histogram} .
The porosity analysis was carried out using the Dragonfly software \cite{dragonfly2024} by following the common procedure outlined in the literature \cite{Koek2024,Desrosiers2024}.
Due to the transparency of $\mathrm{Al_{2}O_{3}}$ to laser radiation, a thin coating of Ni ($\sim100$ nm) was deposited on the front and back surfaces through physical vapor deposition.
When irradiated, the laser beams induce a rapid energy absorption and temperature increase in the Ni coating, which in turn generates a compressive shock wave followed by a release wave that propagates into the $\mathrm{Al_{2}O_{3}}$ substrate.
The interaction of these release waves creates a high hydrostatic tensile state, leading to spall failure if the stress exceeds the spall strength of $\mathrm{Al_{2}O_{3}}$.
Here, the spall strength will be determined experimentally through the spall thickness method \cite{gilath1988laser, Jarmakani2010, Mewael_1st_paper} and compared to high-fidelity interatomic potential under molecular dynamics.
\subsection{Computational setup}
\subsubsection*{Density Functional Theory (DFT) Simulations}
Kohn-Sham density-functional theory (KS-DFT)~\cite{PhysRev.140.A1133,PhysRev.136.B864} was employed through plane-wave-based implementation within the Vienna Ab-Initio Simulation Package (VASP)~\cite{10.1016/0927-0256(96)00008-0}.  
The generalized gradient approximation (GGA) of Perdew, Burke, and Ernzerhof (PBE)~\cite{PhysRevLett.77.3865} was employed as the semilocal exchange-correlation functionals, and the projector augmented-wave (PAW)~\cite{PhysRevB.59.1758} pseudopotentials were used to express the atom core electrons. 
The total energy was minimized for the alumina supercell  (16 unit cells) in all possible degrees of freedom (ionic positions, alumina cell shape, and volume) to determine the optimal relaxed geometry using the self-consistent method where partial wave occupancies were calculated by Gaussian smearing width of $\mathrm{0.026 ~eV}$ as shown in previous study~\cite{10.1021/acs.jpcc.2c06646}. 
The electronic energies cut-off and the kinetic-energy cutoff Ecut are set to $10^{-6}$ eV and 1 meV$\cdot$atom$^{-1}$, respectively. 
A $\mathrm{4 \times 4 \times 4}$ Monkhorst-Pack $k-$points are used as mesh for the alumina crystal to produce the plane wave basis set. 
After fully relaxing the structure, the alumina cell is strained hydrostatically in all directions up to $\pm20\%$ of the lattice parameter with a stepsize of 0.01\% (4000 instances) using AtomProNet python package~\cite{arxiv.org/abs/2501.14039}, and self-consistent field (SCF) calculations are performed to get the geometric features required for Euclidian neural network training.
To include the effect of porosity in the potential energy surface from machine learning potential, four different structures were optimized and hydrostatically strained in DFT simulations, namely- pure alumina, alumina with O vacancy, alumina with Al vacancy, and alumina with both Al, O vacancies.
\subsubsection*{Equivarient neural network interatomic potential}
Equivariant graph neural networks (GNN) based on a 3D Euclidean group have been used to perform symmetry operations~\cite{geiger2022e3nn}. In a graph-based representation of alumina atomic structure, Al and O atoms can be present as the nodes $\left(n_i, n_j\right)$ respectively, and bonds are the edges $\left(e_{i j}\right)$. 
Here, we used Allegro~\cite{Musaelian2023}, an equivariant neural network for a local atomic environment, where scalar and vector features of each atom pass using two-body multi-layer perception ($\phi$). The alumina GNN model was trained with a total allocation of 16000 structures generated through self-consistent DFT calculation, divided into 14000 for training and 2000 for validation. 

All rotations, translations, and reflections symmetry have been preserved in the orthogonal Euclidean group $\mathcal{O}(3)$ symmetry operations.  We used one Allegro layer with one feature of even parity and the hyperparameter that controls maximum rotation, $l_{max}$= 1. The two-body latent multi-layer perceptron (MLP) comprised of three hidden layers of dimensions $\mathrm{[32, 64, 128]}$ using SiLU nonlinearities~\cite{hendrycks2023gaussian} and uniform initialization. The subsequent latent MLP consisted of a single hidden layer with dimension 128, also using a SiLU nonlinearity and uniform initialization, with latent ResNet enabled. The final edge energy MLP had a hidden layer of dimension 32 with uniform initialization and no nonlinearity. Another self-interaction layer reduced the feature dimension to a single scalar output value per atom. The total potential energy was obtained by summing these scalar outputs over all atoms. Forces were derived as the negative gradient of the predicted total potential energy, calculated via automatic differentiation. We used a radial cutoff of 3.0 \text{\AA}. The GNN model was trained using a combined loss function of energies and forces. The loss function weights forces and total energy equally, and the training was performed using the Adam  optimizer~\cite{kingma2017adam} in PyTorch~\cite{NEURIPS2019_bdbca288}. The learning rate and the batch size for this GNN training were 0.002 and 1, respectively. Based on the validation loss, an on-plateau scheduler was employed to decrease the learning rate. The patience was set to 50, and the decay factor was set to 0.5. We employed an exponential moving average, with a weight of 0.99, to assess both the validation set and the final trained model.  Early stopping criteria include a lower bound learning rate of $\mathrm{10^{-5}}$ and patience of 100 epochs for validation loss. The dataset was randomly rearranged after each epoch and trained using the float32 data type. The GNN model was trained on a single NVIDIA V100 GPU.
\subsubsection*{Nanovoid molecular dynamic simulations}
The pre-trained Allegro potential energy surface was used in LAMMPS~\cite{10.1016/j.cpc.2021.108171} to perform classical molecular dynamics simulations.
Unless otherwise stated, the simulation cell dimensions were set to $62a_o \times 60b_o \times 42c_o$ corresponding to a total of $30^3$ unit cells (containing 3,248,624 atoms), where $a_o = 4.85 \, \text{\AA}$, $b_o = 4.99 \, \text{\AA}$, and $c_o = 7.09 \, \text{\AA}$ are the lattice parameters of the $\mathrm{Al_2O_3}$ unit cell \cite{Jain2013, Lin2004, Xu2010}.
An initial void of $d_v$ was created at the center by removing atoms from the bulk $\mathrm{Al_{2}O_{3}}$ super-cell.
Here, as well unless stated, $d_\mathrm{v} = 6$ nm resulting in a void volume fraction ($V_\mathrm{void}/V_\mathrm{cell}$) of 0.4$\%$.
Finally, periodic boundary conditions were applied to the simulation cell to simulate an infinitely large system.
Before loading, the total energy of the $\mathrm{Al_{2}O_{3}}$ super-cell was minimized by iteratively adjusting atom coordinates.
Once minimized, the structure was equilibrated at $T = 300$~K and zero pressure by employing an isothermal-isobaric ensemble for a total duration of 100 ps \cite{martyna1994constant}.

The relaxed structure was then subjected to a controlled deformation gradient, $\textbf{F} = (1 + \dot{\epsilon}\Delta t)\textbf{I}$, to simulate hydrostatic loading, where $\textbf{I}$ is a 3 $\times$3 identify matrix, $\dot{\epsilon}$ is the applied strain rate, and $\Delta t = 5$ fs is the time-step.
Virial stress was computed using the conventional method \cite{allen2017computer} from which the hydrostatic stress $\sigma = (\sigma_{1} + \sigma_{2} + \sigma_{3})/3$ was calculated using the principal stress $\sigma_{i}$.
Volumetric and deviatoric strains were calculated from the atomic Green-Lagrange strain tensor for each atom and averaged with respect to a pre-defined volume to establish a continuum strain measure.
Lastly, OVITO was used to visualize the atomic trajectory and analyze spall failure \cite{stukowski2009visualization}.

\section{Results}
\subsection{Laser spall experiments of alumina}
\begin{figure}[h!]
	\centering
	\begin{tabular}{@{}c@{}}
		\begin{subfigure}[t]{0.3\textwidth}
			\centering
			\includegraphics[width=\linewidth]{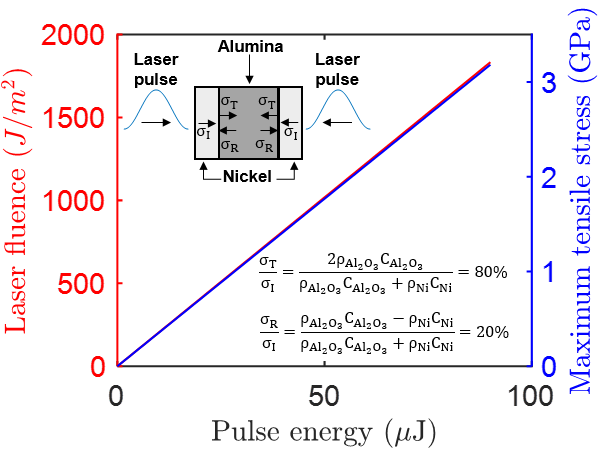}
			\caption{}
			\label{fig:Experiment_panel_1_a}
		\end{subfigure} \\
		\begin{subfigure}[t]{0.3\textwidth}
			\centering
			\includegraphics[width=\linewidth]{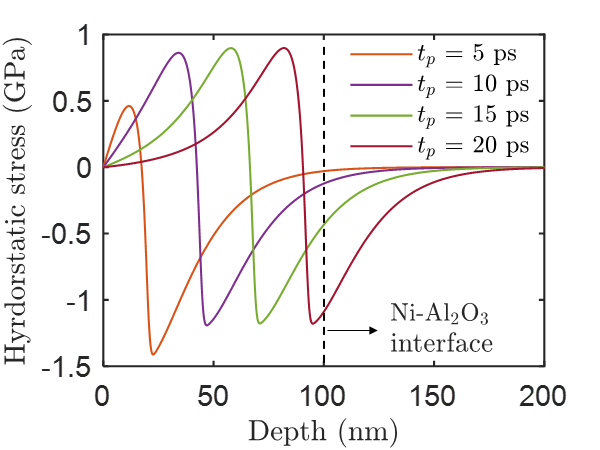}
			\caption{}
			\label{fig:Experiment_panel_1_b}
		\end{subfigure}
	\end{tabular}
	\qquad
	\begin{tabular}{@{}c@{}}
		\textbf{Single-pulse laser approach}\\[0.5em]	    
		\begin{subfigure}[t]{0.45\textwidth}
			\centering
			\includegraphics[width=\linewidth]{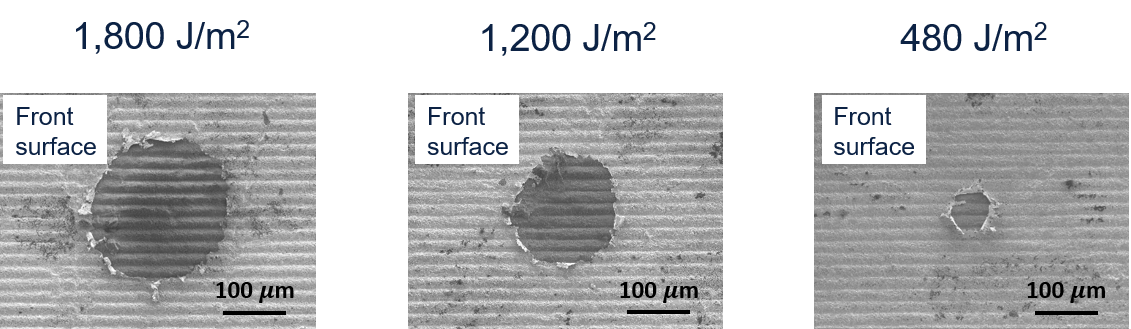}
		\end{subfigure} \\
		\textbf{Two-pulse laser approach}\\[0.5em]	            
		\begin{subfigure}[t]{0.45\textwidth}
			\centering
			\includegraphics[width=\linewidth]{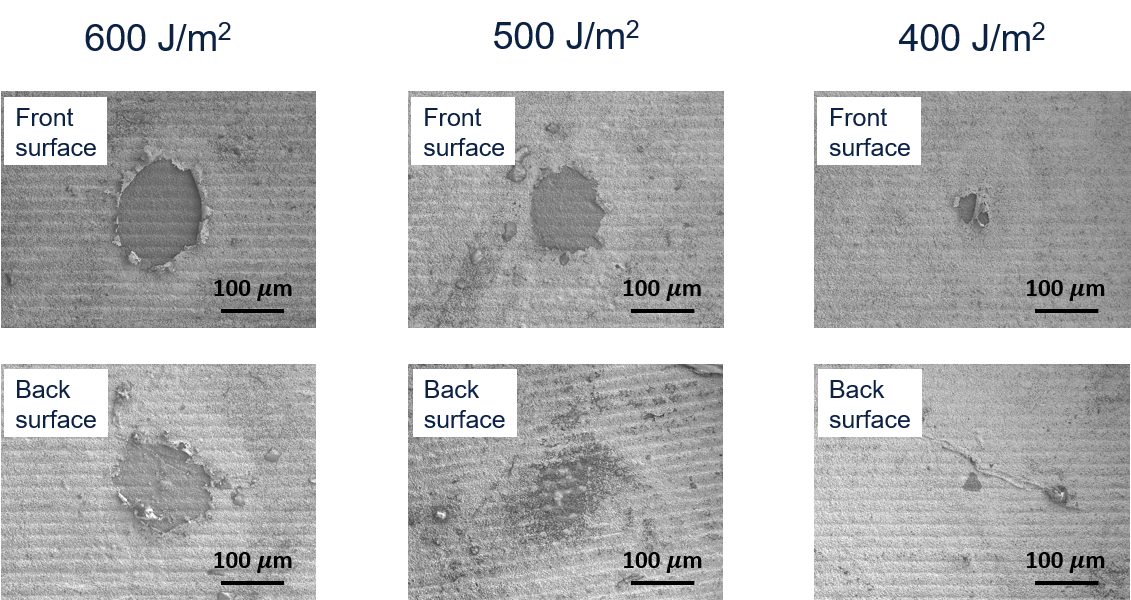}
			\caption{}
			\label{fig:Experiment_panel_1_c}
		\end{subfigure}        
	\end{tabular}
	\caption{Illustration of the laser-induced stress wave propagation and failure mechanisms with analytical calculation of the (a) relationship between pulse energy, laser fluence, and maximum tensile stress for a fixed spot size of $\sim$ 250 $\mu$m and pulse duration of 350 fs, with an inset schematic of the Ni-$\mathrm{Al_{2}O_{3}}$ structure depicting the incident, transmitted, and reflected stress waves and the corresponding ratios for transmitted/incident and reflected/incident stresses, (b) thermoelastic stress wave propagation across the Ni coating as a function of time with a fluence of 600 $\mathrm{J\cdot m^{-2}}$ and a pulse duration of 350 fs, and (c) SEM images of the surface features produced after single- (top) and two-pulse (bottom) laser illumination on the nickel coated $\mathrm{Al_{2}O_{3}}$ samples displaying the front and back surfaces for a range of laser fluence between 1,800 to 400 $\mathrm{J\cdot m^{-2}}$. In these images, the Ni coating (visible as silver toned) is displayed with the underlying $\mathrm{Al_{2}O_{3}}$ substrate exposed with varying degrees depending on the laser fluence.}
\label{fig:Experiment_panel_1}
\end{figure}

Next, we examine the surface response of the nickel coated $\mathrm{Al_{2}O_{3}}$ samples under the single- and two-pulse laser spall approach. 
The stress-strain curves shown in Figure \ref{fig:Molecular_dynamics_panel_b} indicate the amount of hydrostatic tensile stress required to drive spall failure, as a function of the strain rate. 
Using the analytical solution of the two-temperature model \cite{Mewael_1st_paper}, we can relate the laser fluence to the maximum tensile stress developed in the thin Ni coating, as demonstrated in Figure \ref{fig:Experiment_panel_1_a}. 
As seen in the figure, for a fixed spot size of $\sim$ 250 $\mu$m, both the laser fluence and the resulting tensile stress can be controlled by adjusting the pulse energy. 
The incident stress waves ($\sigma_\mathrm{I}$), generated in the Ni coating, propagate towards the $\mathrm{Al_{2}O_{3}}$ substrate. 
Upon reaching the Ni-$\mathrm{Al_{2}O_{3}}$ interface, a portion of the stress wave (80$\%$) is transmitted ($\sigma_\mathrm{T}$) to the center of the sample, while a smaller fraction (20$\%$) is reflected ($\sigma_\mathrm{R}$) back to the free surface.
The ratio of transmitted to reflected stress is governed by the shock impedance of the materials.
The density ($\rho$) of $\mathrm{Al_{2}O_{3}}$ and Ni are 3,890 and 8,900 kg$\cdot$m$^{-3}$, respectively; bulk sound speed (C) of $\mathrm{Al_{2}O_{3}}$ and Ni are 7,455 and 4,850 m$\cdot$s$^{-1}$, respectively \cite{Murray1998,Reinhart2003}.
Prior to reaching the interface, following laser-matter interaction, the Ni coating experiences a rapid increase in temperature and pressure, and propagates shockwaves towards the $\mathrm{Al_{2}O_{3}}$ substrate, containing a compression front followed by an unloading tensile wave, as shown in Figure \ref{fig:Experiment_panel_1_b}.
It should be noted that the stress waves calculated in Figures. \ref{fig:Experiment_panel_1_a} and \ref{fig:Experiment_panel_1_b} account for Ni reflectivity (77$\%$) \cite{Johnson1974}.
The passing of these unloading tensile waves across the interface could drive interface failure, and as these waves propagate toward the bulk, their interaction could also drive spall failure.
Here, Figure \ref{fig:Experiment_panel_1_b} illustrates the effects of a single-pulse; in the case of the two-pulse approach, another set of stress waves would be generated at the opposite surface.

Figure \ref{fig:Experiment_panel_1_c} displays the surface observation of the Ni coated $\mathrm{Al_{2}O_{3}}$ samples under the single- and two-pulse laser approach for a range of laser fluences.
Here, for a fixed spot size of $\sim$ 250 $\mu$m, the laser fluence is adjusted by varying the pulse energy through a power attenuator.
Once the sample was irradiated above a certain fluence threshold, sufficient tensile stress for interface separation was generated to induce Ni-$\mathrm{Al_{2}O_{3}}$ adhesion failure due to the nucleation, growth, and coalescence of voids at the interface.
Under SEM imagining, it can be seen that varying the laser fluence from 1,800 to 400 $\mathrm{J\cdot m^{-2}}$ led to a reduction in the surface damage region, i.e., crater, from 230 $\mu$m to 56 $\mu$m, respectively--due to the reduction in tensile stress with decreasing laser fluence (see Figure \ref{fig:Experiment_panel_1_b}).
The ejection of the Ni layer after interface debonding is most likely due to the outward propagation of the tensile stress waves and the radial compressive waves \cite{deRessguier2021}.
In the conventional single-pulse laser approach, interface failure takes place at the front surface only, however, introducing the secondary laser pulse led to failure at both the front and back surfaces.
However, in both approaches, the surface features are similar highlighted with dangling Ni flakes and micro-voids on the Ni surface.
Close inspection (as shown in Figure \ref{fig:surface_observation_close_up_laser_fluence_1800_and_400_J_m2}) of the surface morphology suggests that fracture under the dynamic tensile stress occurred mainly on the interface rather than on the Ni film indicating that the local adherence between Ni and $\mathrm{Al_{2}O_{3}}$ might be weaker than the spall strength of Ni.

\begin{figure}[h!]
\centering
\begin{subfigure}[b]{0.32 \textwidth}
	\centering
	\includegraphics[width=\linewidth]{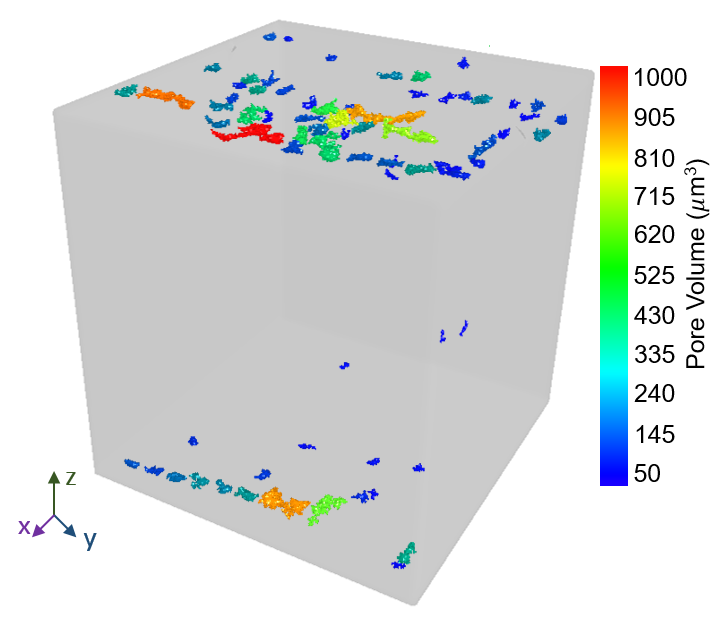}
	\caption{}
	\label{fig:Experiment_panel_2_a}
\end{subfigure}
\hspace{2.0cm}
\begin{subfigure}[b]{0.32 \textwidth}
	\centering
	\includegraphics[width=\linewidth]{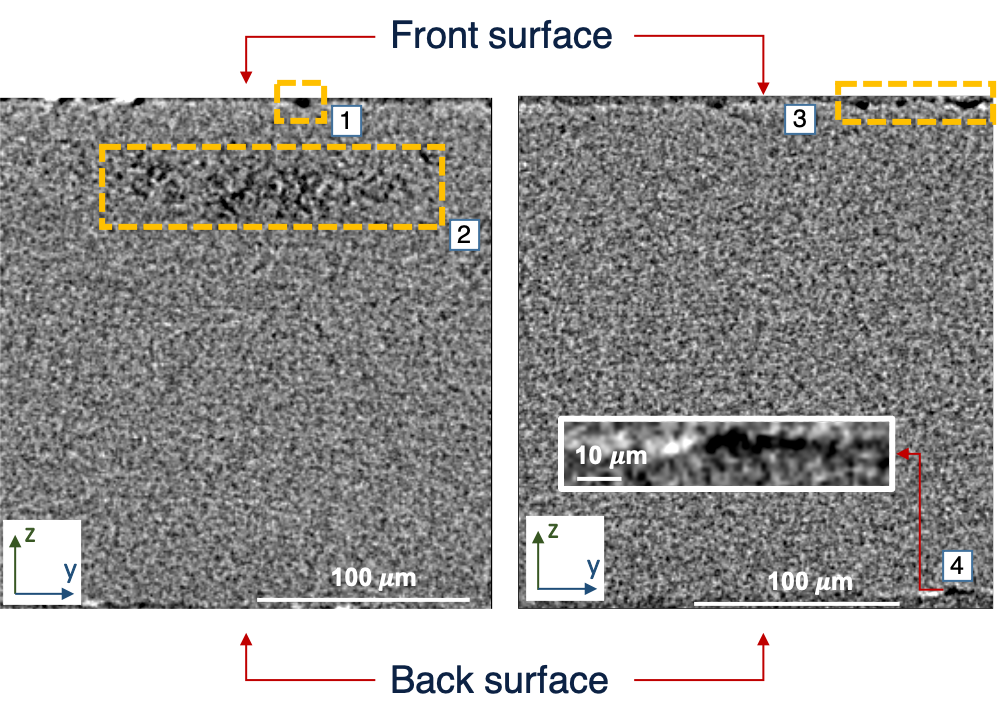}
	\caption{}
	\label{fig:Experiment_panel_2_b}
\end{subfigure}	
\par
\begin{subfigure}[b]{0.32 \textwidth}
	\centering
	\includegraphics[width=\linewidth]{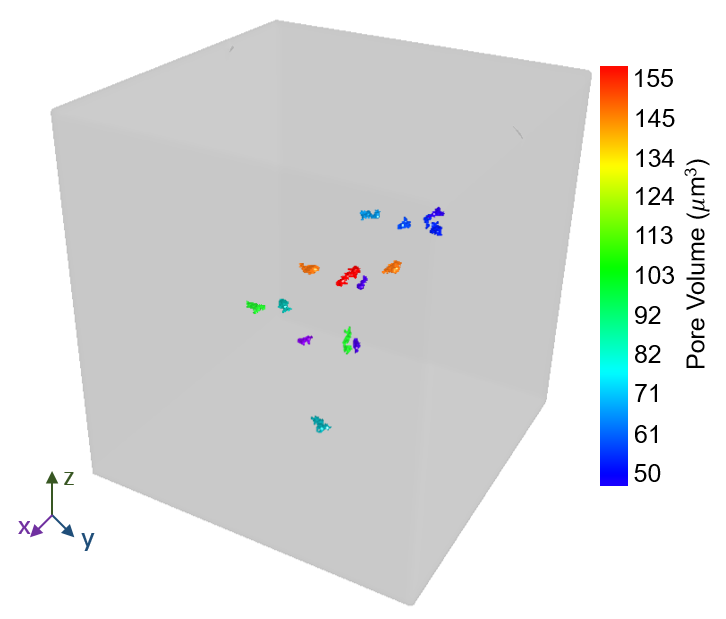}
	\caption{}
	\label{fig:Experiment_panel_2_c}
\end{subfigure}
\hspace{2.0cm}
\begin{subfigure}[b]{0.32 \textwidth}
	\centering
	\includegraphics[width=\linewidth]{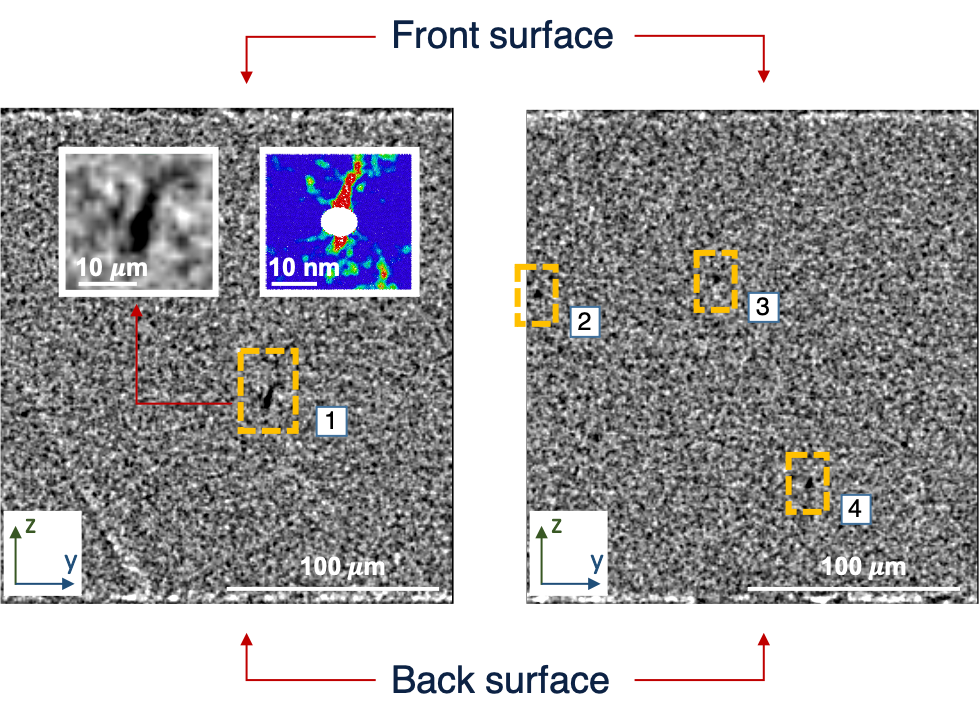}
	\caption{}
	\label{fig:Experiment_panel_2_d}
\end{subfigure}
\par
\begin{subfigure}[b]{0.32 \textwidth}
	\centering
	\includegraphics[width=\linewidth]{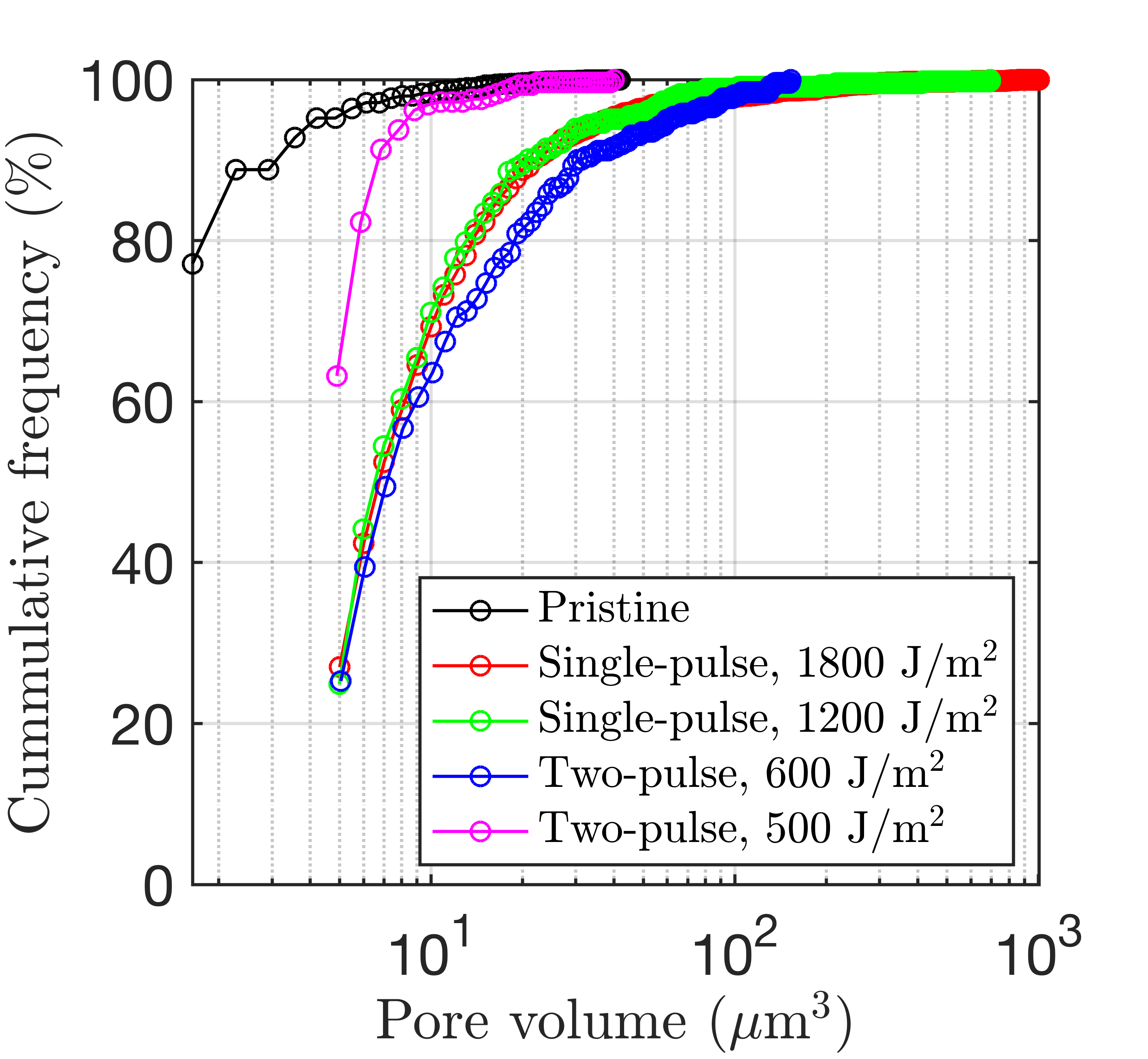}
	\caption{}
	\label{fig:Experiment_panel_2_e}
\end{subfigure}	
\hspace{2.0cm}
\begin{subfigure}[b]{0.32 \textwidth}
	\centering
	\includegraphics[width=0.95\linewidth]{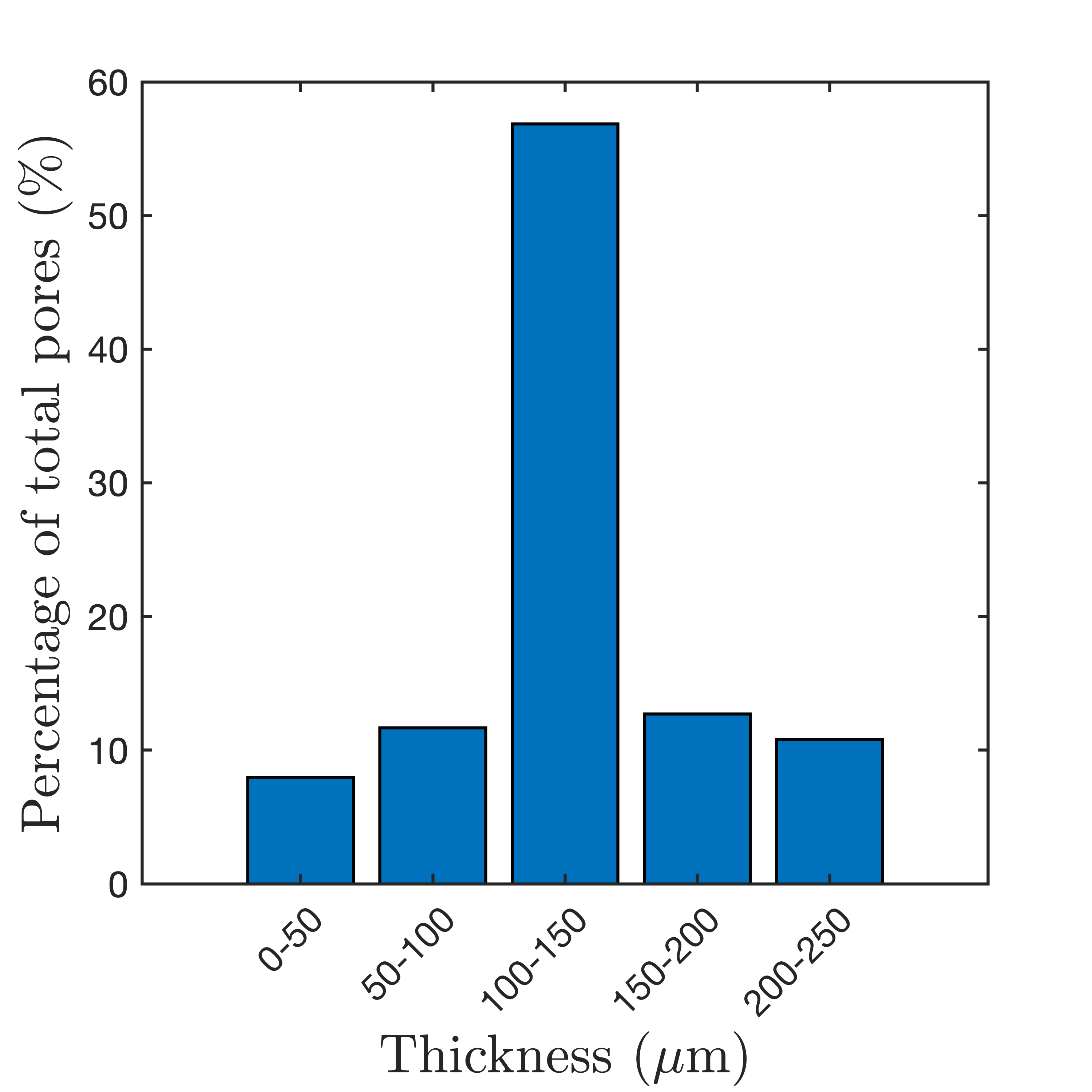}
	\caption{}
	\label{fig:Experiment_panel_2_f}
\end{subfigure}
\caption{Illustration of spall failure in $\mathrm{Al_{2}O_{3}}$ induced by the (a,b) single- (fluence: 1,800 $\mathrm{J\cdot m^{-2}}$) and (c,d) two-pulse (fluence: 600 $\mathrm{J\cdot m^{-2}}$) laser approaches, where (a, c) illustrates the 3D volume reconstructions from micro-CT scans with blue and red voxels indicating smaller and larger voids, (b, d) presents the 2D slices, also from micro-CT scans, highlighting failure locations associated with each approach, (e) displays the cumulative pore volume distribution comparing pristine and post-loaded samples under the single- (1,800 and 1,200 $\mathrm{J\cdot m^{-2}}$) and two-pulse (600 and 500 $\mathrm{J\cdot m^{-2}}$) laser approaches, and (f) describes the pore distribution for the two-laser setup at 600 $\mathrm{J\cdot m^{-2}}$. The inset figure in (b) highlights a zoomed-in image of the largest void closest to the back surface, while (d) highlights a zoomed-in image of a central void and compares the crack morphology to that in Figure \ref{fig:Molecular_dynamics_panel_d} at 5$\%$ strain.}
\label{fig:Experiment_panel_2}
\end{figure}

To investigate the effects of single- and two-pulse laser approaches on the spall behavior of bulk $\mathrm{Al_{2}O_{3}}$, micro-CT analysis was conducted on two cases: a single-pulse laser with a fluence of 1,800 $\mathrm{J\cdot m^{-2}}$ and a two-pulse laser with a reduced fluence of 600 $\mathrm{J\cdot m^{-2}}$, as shown in Figs. \ref{fig:Experiment_panel_2}a---d.
A 3D pore structure was reconstructed using standard micro-CT segmentation and thresholding techniques \cite{Koek2024,Desrosiers2024}, where blue voxels represent smaller voids and red voxels indicate larger ones. In Figs. \ref{fig:Experiment_panel_2_a} and \ref{fig:Experiment_panel_2_c}, only pores with a volume larger than 50 $\mathrm{\mu m}^3$ are shown for better visualization.
In the single-pulse approach (fluence: 1,800 $\mathrm{J\cdot m^{-2}}$), numerous voids were observed across the sample, with the larger ones concentrated near the front and back surfaces.
In contrast, the two-pulse approach (fluence: 600 $\mathrm{J\cdot m^{-2}}$) resulted in fewer voids, with the larger ones predominantly located near the sample's center. The lower laser fluence in this approach led to fewer large voids being nucleated, as the generated tensile stress was lower (c.f., Figs. \ref{fig:Experiment_panel_2_a} and \ref{fig:Experiment_panel_2_c}).
However, despite the lower number of voids, incipient spall failure still occurred in the two-pulse approach, as the interaction of unloading tensile waves enabled failure even at reduced fluence.

To further analyze the spall location and mechanisms, 2D slices of the samples were examined, as shown in Figs. \ref{fig:Experiment_panel_2_b} and \ref{fig:Experiment_panel_2_d}.
For the single-pulse laser approach, clusters of voids were observed near the front and back surfaces.
These voids formed primarily due to the passage of the unloading tensile wave (locations 1, 2, and 3 in Figure \ref{fig:Experiment_panel_2_b}) and the interaction of the reflected compressive wave with the unloading wave near the back surface (location 4 in Figure \ref{fig:Experiment_panel_2_b}). 
Additionally, pre-existing defects, wave reverberations, and partial reflections may have further contributed to the void distribution.
In contrast, the two-pulse laser approach primarily induced central voids (locations 1, 2, 3, and 4 in Figure \ref{fig:Experiment_panel_2_d}). 
These voids were likely caused by the interaction of the two unloading tensile waves, which is characteristic of this method \cite{Mewael_1st_paper}. 
By incorporating a secondary pulse, the two-pulse approach shifts the spall location toward the sample center, as observed in Figure \ref{fig:Experiment_panel_2_d}.
%
%
Furthermore, upon further analysis of the void morphology of the central void at location in one with the high-fidelity MD simulation in the inset figures in Figure \ref{fig:Experiment_panel_2_d}, it is clear to see that similarities can be drawn.
The micro-CT scan reveals a crack pattern indicating some degree of propagation in an angled direction, suggesting an anisotropic effect similar to that observed in the MD simulations.
Although crack initiation was not observed in the experiments, it can be inferred that the micro-crack propagated from an initial pore under a state of hydrostatic tensile stress ($\sim1.8$~GPa according to Figure \ref{fig:Experiment_panel_1_b}).
This qualitative observation thus demonstrates that the macroscopic failure observed experimentally can be linked to the atomic-scale nano-void failure seen in the MD simulations.

Porosity analysis of the pristine sample (see Figure \ref{fig:pristine_histogram}) showed that 95$\%$ of pores had a volume below 3.58 $\mu$m$^{3}$.
After applying a size threshold to remove smaller pores, the pore volume distribution of the post-load samples is shown in Figure \ref{fig:Experiment_panel_2_e}.
For single-pulse cases at fluences of 1,200 and 1,800 $\mathrm{J\cdot m^{-2}}$), a similar trend was observed where the smallest and largest pore size were almost identical.  
%
%
Let us now focus on the two-pulse experiment also shown in Figure \ref{fig:Experiment_panel_2_e}.  
At 500 $\mathrm{J\cdot m^{-2}}$, no spall damage was observed, as the pore distribution remained similar to the pristine case with differences in the small pore sizes. 
Increasing the fluence to 600 $\mathrm{J\cdot m^{-2}}$, we observed that the pore distribution matches quite neatly the single pulse experiments at 1200 and 1800 $\mathrm{J\cdot m^{-2}}$. 
Noteworthy, the size of the largest pores is smaller for the two-laser setup due to the lower fluence of the single pulse (600 $\mathrm{J\cdot m^{-2}}$) and the decay effects of the shock pressure. 
Thus, the spall threshold for the two-pulse approach is estimated to be between 500 and 600 $\mathrm{J\cdot m^{-2}}$.
Comparing the single- and two-pulse cases at 1,800 and 600 $\mathrm{J\cdot m^{-2}}$, respectively, the single-pulse case generated more numerous smaller voids compared to the two-pulse case but resulted in a lower 95th percentile pore volume (38.1 $\mu$m$^{3}$ vs. 59.8 $\mu$m$^{3}$).
However, the single-pulse approach produced a larger maximum pore volume (1,000 $\mu$m$^{3}$ vs. 150 $\mu$m$^{3}$) because of the higher hydrostatic tensile stress.

Figure \ref{fig:Experiment_panel_2_f} shows the distribution of pore across the sample's thickness for the two-laser setup at 600 $\mathrm{J\cdot m^{-2}}$. Remarkably, we can observe that the pore location is concentrated near the center of the sample, where the interaction between release waves happened. 
In comparison, the pore distribution observed with the single-pulse setup (see Figure \ref{fig:pore_distribution_panel_single_two_pulse}) reveals that, due to wave release and reflection, the spall planes appeared near the free surface of the samples, which is typical of the laser spall experiment. 
Moreover, pore distribution analysis (see Figure \ref{fig:pore_distribution_panel_single_two_pulse}) showed two spall planes at 1,800 $\mathrm{J\cdot m^{-2}}$, whereas only one spall plane near the illuminated surface at 1,200 $\mathrm{J\cdot m^{-2}}$. 
This difference is due to the magnitude of the stress generated by the different fluence and the decay of the pressure with thickness.
This highlights the two-pulse approach's ability to induce spall failure at a specific location of the sample through the thickness, opening up potential applications to understand the spall failure of interfaces better. 

\subsection{Potential development and validation}
\begin{figure}[ht]
\centering
\begin{subfigure}[b]{0.6\textwidth}
	\centering
	\includegraphics[width=\linewidth]{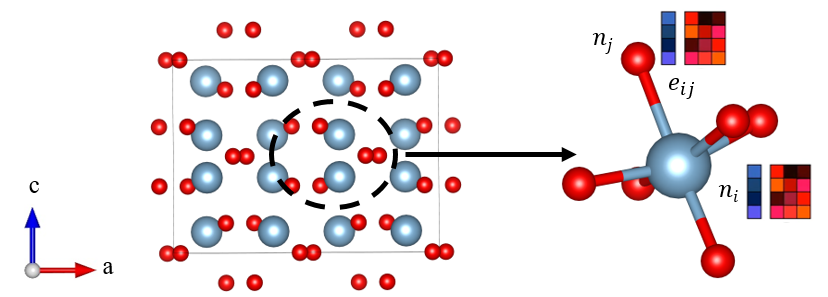}
	\caption{}
	\label{fig:Schematic_Parity_Plot_a}
\end{subfigure}
\hfill
\begin{subfigure}[b]{0.3\textwidth}
	\centering
	\includegraphics[width=\linewidth]{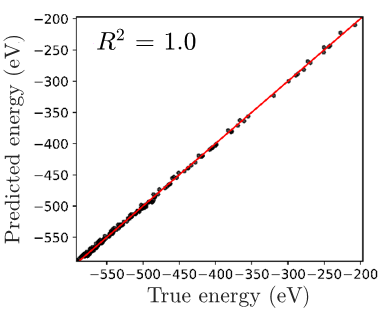}
	\caption{}
	\label{fig:Schematic_Parity_Plot_b}
\end{subfigure}
\hfill
\begin{subfigure}[b]{0.3\textwidth}
	\centering
	\includegraphics[width=\linewidth]{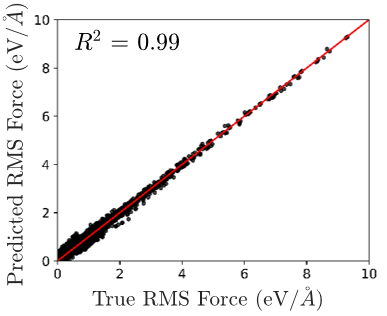}
	\caption{}
	\label{fig:Schematic_Parity_Plot_c}
\end{subfigure}	
\hfill
\begin{subfigure}[b]{0.3\textwidth}
	\centering
	\includegraphics[width=\linewidth]{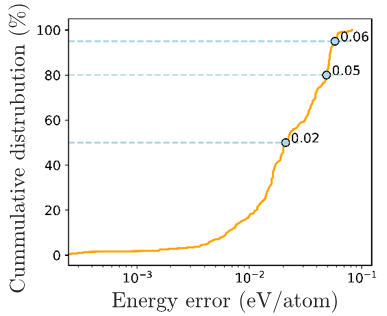}
	\caption{}
	\label{fig:Schematic_Parity_Plot_d}
\end{subfigure}		
\hfill
\begin{subfigure}[b]{0.3\textwidth}
	\centering
	\includegraphics[width=\linewidth]{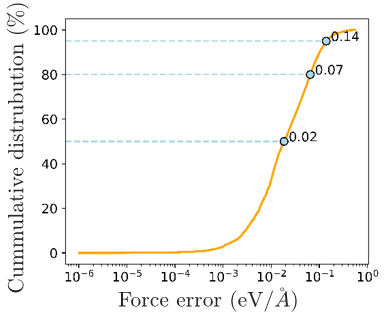}
	\caption{}
	\label{fig:Schematic_Parity_Plot_e}
\end{subfigure}			
\caption{(a) Node (Al, O atoms) and edge (Al-O bond) update using the graph neural network. Here, the color codes for alumina atoms are red-oxygen and blue- aluminum. Here, the parity plots compare the Allegro-trained (predicted) model with DFT (true) for (b) energy and (c) root mean square (RMS) forces along with $\mathrm{R^2}$ values; the cumulative distribution plots compare the absolute error (\%) between the Allegro-trained model and DFT-relaxed alumina crystal in (d) energies and (e) forces.}
\label{fig:Schematic_Parity_Plot}
\end{figure}
Having established experimental insights into the spall behavior of Al$_2$O$_3$, we now turn to molecular dynamics simulations to further investigate the material's response at \textbf{ps} timescales and \textbf{nm} length scales.
To enable accurate atomistic simulations, we developed an interatomic potential using the Allegro framework and first validate its accuracy against density functional theory (DFT) results.
Figure \ref{fig:Schematic_Parity_Plot_a} depicts the crystal structure of Al$_2$O$_3$ with atoms as spheres (nodes of the graph, i.e., $n_i$) and bonds as cylinders (edges of the graph, i.e., $e_{ij}$).  
Figures \ref{fig:Schematic_Parity_Plot_b}–\ref{fig:Schematic_Parity_Plot_c} show parity plots of the validation dataset, demonstrating strong agreement between the DFT ground truth and the trained model predictions, as reflected by the high $\mathrm{R^2}$ values (close to 1) for energy and force predictions. 
Here, the energy refers to the total potential energy of the alumina supercell, while the force represents the per-atom force acting on each atom.
The cumulative distribution of the Allegro model errors, depicted in Figures \ref{fig:Schematic_Parity_Plot_d}---\ref{fig:Schematic_Parity_Plot_e}, demonstrates the model's accuracy, with horizontal dashed lines marking the 50th, 80th, and 95th percentiles of the energy and force errors.
From these distributions, 50\% of the data exhibit energy and force errors below 0.02 eV per atom and  0.02 $\mathrm{eV\cdot\text{\AA}^{-2}}$; 80\% fall below 0.05 eV per atom and  0.07 $\mathrm{eV\cdot \text{\AA}^{-2}}$; and 95\% remain below 0.06 eV per atom and  0.14 $\mathrm{eV\cdot \text{\AA}^{-2}}$. 
The parity and cumulative plots indicate that the Allegro model predictions closely match DFT results, particularly given the stringent DFT training criteria(energy convergence: 0.001 meV, force convergence: 1 meV$\cdot$\text{\AA}$^{-1}$). 
Thus, these results demonstrate that the Allegro model achieves near-DFT-level accuracy, validating its reliability for atomistic simulations.
After validating the Allegro potential against DFT, we further assessed its accuracy by comparing its predictions for key material properties of single-crystal alumina (elastic constants, cohesive energy, and vacancy formation energy) with those from two widely used interatomic potentials, COMB3 and ReaxFF.
These values were also compared with DFT computed values available in the literature \cite{deJong2015}.
Table \ref{tb:Elastic_constant_comparison} highlights the accuracy of the Allegro potential in comparison to the selected reactive potentials.
It should be noted that for Allegro potential, the predicted bulk modulus, calculated classically with deviatoric information \cite{reuss1929berechnung,voigt1928lehrbuch}, is higher than the DFT and COMB3 calculated values but remarkably lower than ReaxFF. 
\begin{table}[h!]
\centering
\caption{Comparison of elastic constants, bulk modulus ($K$), cohesive and vacancy formation energy among DFT \cite{deJong2015}, Allegro-MD, COMB3-MD , and ReaxFF-MD  at ground state condition.}
{\begin{tabular}{lcccc}
		\hline
		\hline
		& DFT & Allegro-MD & COMB3-MD & ReaxFF-MD\\ 
		\hline     
		\hline
		C\(_{11}\) (GPa) & 454 & 471 & 650 & 853 \\
		C\(_{22}\) (GPa) & 394 & 376 & 472 & 878 \\
		C\(_{33}\) (GPa) & 466 & 473 & 565 & 776 \\
		$K$ (GPa) & 232 & 290 & 270 & 508 \\
		$E_\mathrm{coh}$ (eV/atom) & -7.397 & -7.398 & -6.099 & -6.014 \\
		O vacancy (eV/atom) & -7.343 & -7.347 & -6.110 & -5.949 \\
		Al vacancy (eV/atom) & -7.296 & -7.384 & -6.037 & -5.950 \\
		Al-O vacancy (eV/atom) & -7.312 & -7.378 & -6.052 & -5.933 \\  
		\hline
		\hline
\end{tabular}}
\label{tb:Elastic_constant_comparison}
\end{table}

\subsection{Nanovoid Simulations}

\begin{figure}[ht]
\centering
\begin{subfigure}[b]{0.32\textwidth}
	\centering
	\includegraphics[width=\linewidth]{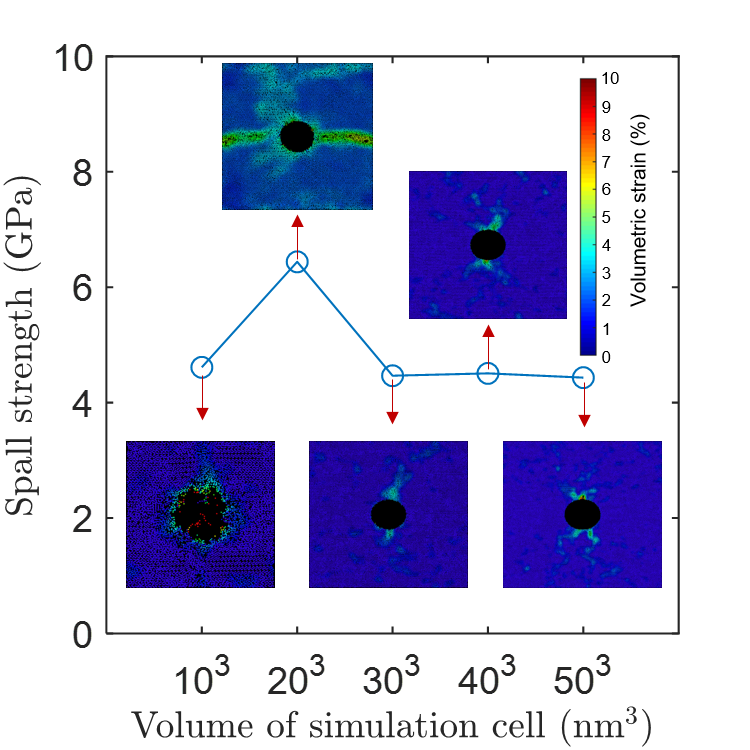}
	\caption{}
	\label{fig:Molecular_dynamics_panel_a}
\end{subfigure}
\hfill
\begin{subfigure}[b]{0.32\textwidth}
	\centering
	\includegraphics[width=\linewidth]{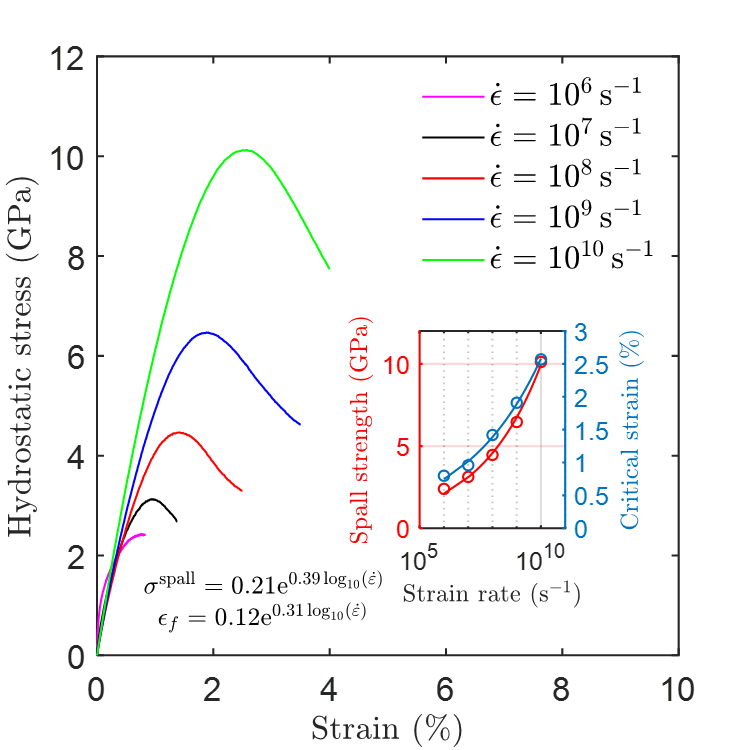}
	\caption{}
	\label{fig:Molecular_dynamics_panel_b}
\end{subfigure}
\hfill
\begin{subfigure}[b]{0.32\textwidth}
	\centering
	\includegraphics[width=\linewidth]{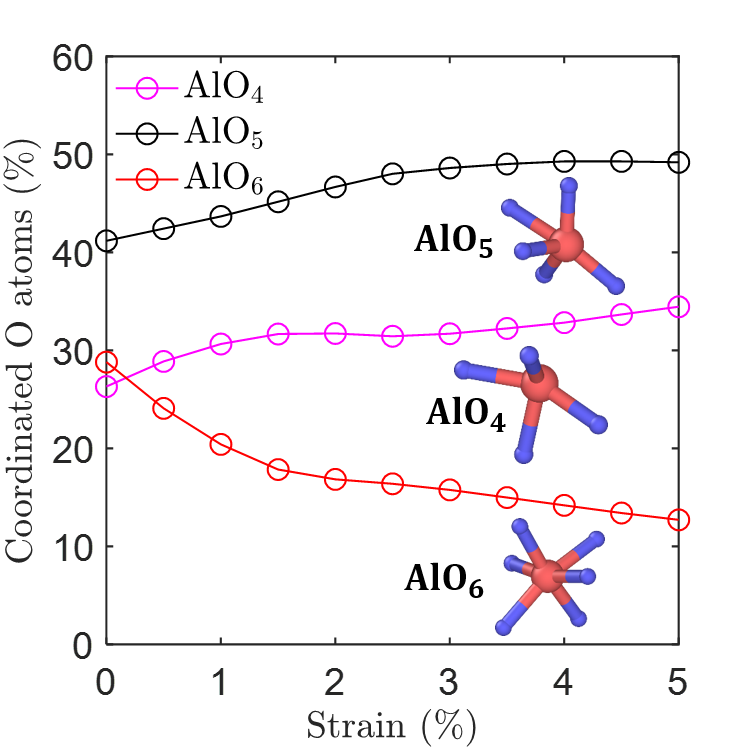}
	\caption{}
	\label{fig:Molecular_dynamics_panel_c}
\end{subfigure}	
\hfill
\begin{subfigure}[b]{0.9\textwidth}
	\centering
	\includegraphics[width=\linewidth]{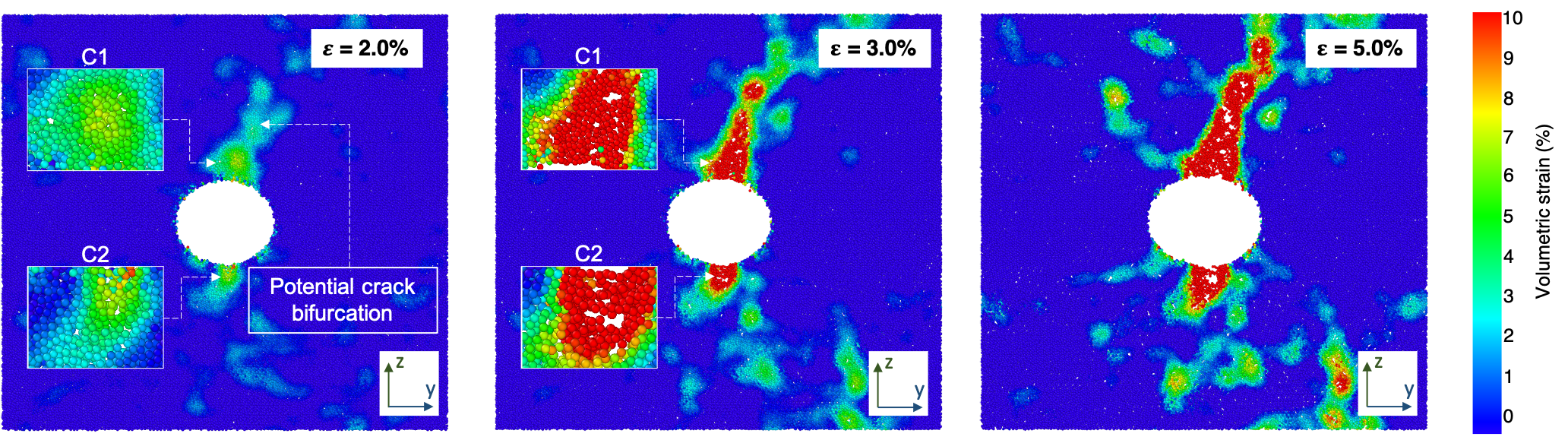}
	\caption{}
	\label{fig:Molecular_dynamics_panel_d}
\end{subfigure}		
\caption{Molecular dynamics simulation of $\mathrm{Al_{2}O_{3}}$ under hydrostatic loading with the following figures highlighting the (a) spall strength of five different simulation cells with a nanovoid corresponding to a void fraction 0.4$\%$ and the microstructure of each cell at 2$\%$ strain, (b) hydrostatic stress for different strain rates with an inset figure of the relationship between the spall strength, $\sigma_{\text{spall}}$, and critical strain, $\epsilon_f$, with the strain rate, (c) change in coordination number of Al atoms with increasing strain and (d) spall behavior under a strain rate of $\mathrm{10^{8}~s^{-1}}$ for a simulation cell with a volume of 30 $\mathrm{nm^{3}}$ and a nano-void diameter of 6 nm.}
\label{fig:Molecular_dynamics_panel}
\end{figure}

To better illustrate the need for more accurate Al$_2$O$_3$ potentials, we provide a stress-strain and failure comparison between of Allegro, COMB3, and ReaxFF in Figure \ref{fig:Molecular_dynamics_COMB_ALLEGRO_REAXFF_comparison}.  
From the comparison, it is evident that the stress values obtained from COMB3 are overpredicted, whereas the results from Allegro and ReaxFF are comparable.
However, while Allegro predicts a birttle type of failure -as explained later-, ReaxFF predicts a ductile failure mode with phase transformation of the material. 
Let us now describe the behavior obtained with the Allegro potential. 
The initial phase of our analysis on the behavior of $\mathrm{Al_{2}O_{3}}$ focuses on the relationship between the simulation cell volume and spall strength at strain rate of $\mathrm{10^{8}~s^{-1}}$. 
It is well established from the literature that cell size plays an impact on the accuracy of predicted physics \cite{Ponga2016, Grgoire2017}.
To explore this, simulations were conducted across five different cell volumes: $V_\mathrm{cell}$ = 10 $\mathrm{nm^{3}}$ (128,398 atoms), 20 $\mathrm{nm^{3}}$ (994,864 atoms), 30 $\mathrm{nm^{3}}$ (3,248,624 atoms), 40 $\mathrm{nm^{3}}$ (7,660,318 atoms), and 50 $\mathrm{nm^{3}}$ (14,911,784 atoms). 
A fixed void volume of 0.4$\%$ was maintained across all cases to ensure consistency.
The results, illustrated in Figure \ref{fig:Molecular_dynamics_panel_a}, show that the spall strength becomes relatively insensitive to simulation cell size when $V_\mathrm{cell}$ reaches approximately 30 $\mathrm{nm^{3}}$.
Additionally, the microstructural analysis, at 2$\%$ strain, in Figure \ref{fig:Molecular_dynamics_panel_a} reveals that the 20 $\mathrm{nm^{3}}$ cell size predicted a failure unlike the initial void deformation observed with other cell sizes. 
These findings suggest that a cell size of 30 $\mathrm{nm^{3}}$ effectively minimizes periodic boundary effects during hydrostatic loading simulations.
Therefore, we selected this cell size for subsequent investigations into the strain rate's influence on $\mathrm{Al_{2}O_{3}}$ behavior and to further analyze the failure mechanisms that lead to spallation.
Next, we proceed to describe the evolution of hydrostatic stress with strain under hydrostatic loading conditions for strain rates between $\mathrm{10^{6}~s^{-1}}$ to $\mathrm{10^{10}~s^{-1}}$.
As shown in Figure \ref{fig:Molecular_dynamics_panel_b}, the material initially undergoes a linear elastic response for all strain rates until reaching a critical volumetric strain, $\epsilon_{f}$. 
Beyond this point, the material exhibits spall failure with anisotropic behavior, as evidenced from the calculated elastic constants in Table \ref{tb:Elastic_constant_comparison}.
As expected, the stress values were more pronounced along the \textbf{Y} direction (corresponding to the \textbf{22} direction), however, in all cases, brittle type of failure was consistently observed despite anisotropic effects \cite{Zhang2008,Nishimura2008}.
Here, the spall strength, i.e., the resistance of a material to fail under nanovoid growth, is characterized as the stress peak in Figure \ref{fig:Molecular_dynamics_panel_b}.
Both the spall strength and the critical strain exhibit a strong strain rate dependency, well-characterized by an exponential relationship. 
Specifically, the spall strength $\sigma_\mathrm{spall}$ can be described by an exponential model of the form $\sigma_\mathrm{spall} = m \times e^{k\dot{\epsilon}}$, where $m$ and $k$ are fitting parameters, indicating that higher strain rates lead to significantly increased spall strength. 
For example, spall strength increases from 2.4~GPa at $10^6~\mathrm{s^{-1}}$ to 10.1~GPa at $10^{10}~\mathrm{s^{-1}}$.
Similarly, the critical strain increases from 0.8\% to 2.6\% over the same range of strain rates.
The stress-strain relationship demonstrates that the energy required to induce failure in $\mathrm{Al_{2}O_{3}}$ increases with the strain rate. 
The energy required to cause failure rises from $14.5~\mathrm{MJ \cdot m^{-3}}$ at $10^6~\mathrm{s^{-1}}$ to $170.5~\mathrm{MJ\cdot m^{-3}}$ at $10^{10}~\mathrm{s^{-1}}$, highlighting the material's increased resistance to failure under rapid deformation conditions.
This increased resistance can be attributed to the inhibition of relaxation processes, such as bond breakage and crack propagation at higher rates, resulting in accumulated internal stresses.
We proceed to explain in detail the spall failure behavior of $\mathrm{Al_{2}O_{3}}$ under hyrdrostatic loading.
It should be noted that despite the differences in spall strength and critical strain, the failure mechanism is invariant with respect to the strain rate.
As such, the spall behavior discussed here corresponds to $10^{8}~\mathrm{s^{-1}}$.
Prior to reaching the critical strain ($= 1.4 \%$), the Al-O bonds are stretched elastically without failure.
However, upon reaching it, bond breakage starts to take place very close to the nano-void as shown in Figure \ref{fig:Molecular_dynamics_bond_breakage}.
Due to the presence of the nanovoid, stress concentration exists leading to larger stress states closer to the nanovoid which is the prime site for failure initiation as seen here, and reported in the literature \cite{Jayatilaka1977,Griffith1921,Bavdekar2022}.
An influential factor indicating bond breakage is the coordination number, which corresponds to the number of bonds per each Al and O atom.
Here, we focus on the Al-O bonds, as they were far more in number in comparison to Al-Al and O-O bonds.
Figure \ref{fig:Molecular_dynamics_panel_c} illustrates the change in the coordination number of Al atoms, where we track 4-coordinated ($\mathrm{AlO_{4}}$), 5-coordinated ($\mathrm{AlO_{5}}$), and 6-coordinated ($\mathrm{AlO_{6}}$) Al atoms with increasing strain. 
Here, it can be seen that with increasing strain the total number of $\mathrm{AlO_{6}}$ bonds reduces to $\mathrm{AlO_{4}}$ and $\mathrm{AlO_{5}}$ indicating an increase in bond breakage.
Hence with the increase in tension beyond the critical strain, Al-O interatomic bonds tend to cleave leading to initial brittle spall failure.
The coalescence of these broken bonds form micro-cracks, and this can be seen in Figure \ref{fig:Molecular_dynamics_panel_d}, when the strain reaches $2\%$.
Figure \ref{fig:Molecular_dynamics_panel_d} illustrates a series of 2D slices, along the \textbf{YZ} plane, of the brittle spall failure originating from the nanovoid.
Here, we can see the formation of two localized regions (highlighted by the increased volumetric strain) close to the nanovoid where micro-cracks formation takes place (labeled as C1 and C2).
In the figure, the red regions represent atoms that experienced $\epsilon \rightarrow 10.0\%$, and blue atoms $\epsilon \rightarrow 0.0\%$.
These localized regions, C1 and C2, are created as a result of the increased displacement of atoms due to stress concentration effects.
However, it is evident that some degree of anisotropy took place leading to specific orientation of these microcracks at an angle from the nanovoid.
Furthermore, these regions are also characterized with vacancies arising due to the bond breakage (see Figure \ref{fig:Molecular_dynamics_bond_breakage}).
It can be seen that when $\epsilon = 3.0\%$, with the increase in tension, atoms in C1 and C2 experience large volumetric strain leading to the nucleation of voids.
Interestingly, when comparing the microstructure at $\epsilon = 2.0\%$ and $3.0\%$ that C1, instead of bifurcating, expanded into a larger crack as it was more energetically favorable to do so.
Lastly, at $\epsilon = 5.0\%$, it can be seen that while C2 expercienced a crack arrest, C1 propagates to the edges of the periodic simulation cell.
However, due to the anistropic behavior of $\mathrm{Al_{2}O_{3}}$, the failure behavior is dependent on the orientation (cf., Figure \ref{fig:Molecular_dynamics_failure_XY_XZ_plane}).
Overall, it can bee seen that the governing mechanisms behind the brittle failure of $\mathrm{Al_{2}O_{3}}$ under high-strain-rate hydrostatic loading are bond breakage, micro-crack formation and coalescence, crack propagation, and crack bifurcation.
As these cracks grow and propagate, they eventually coalescence to generate a free surface as shown in Figure \ref{fig:Molecular_dynamics_failure_surf_mesh}.
It is evident that the initial nanovoid acted as the prime site for failure, as multiple microcracks formed close to the vicinity to then expand and form a spall plane.
These findings compare very well with existing experimental findings \cite{Louro1989,Bourne2001,Antoun2003}.
Furthermore, the nanovoid size impacts both the degree of failure and spall strength as can be found in Figure \ref{fig:Molecular_dynamics_void_size_strain_rate}.

\subsection{Spall strength of alumina}
\begin{figure}[ht]
\centering
\includegraphics[width=0.75\linewidth]{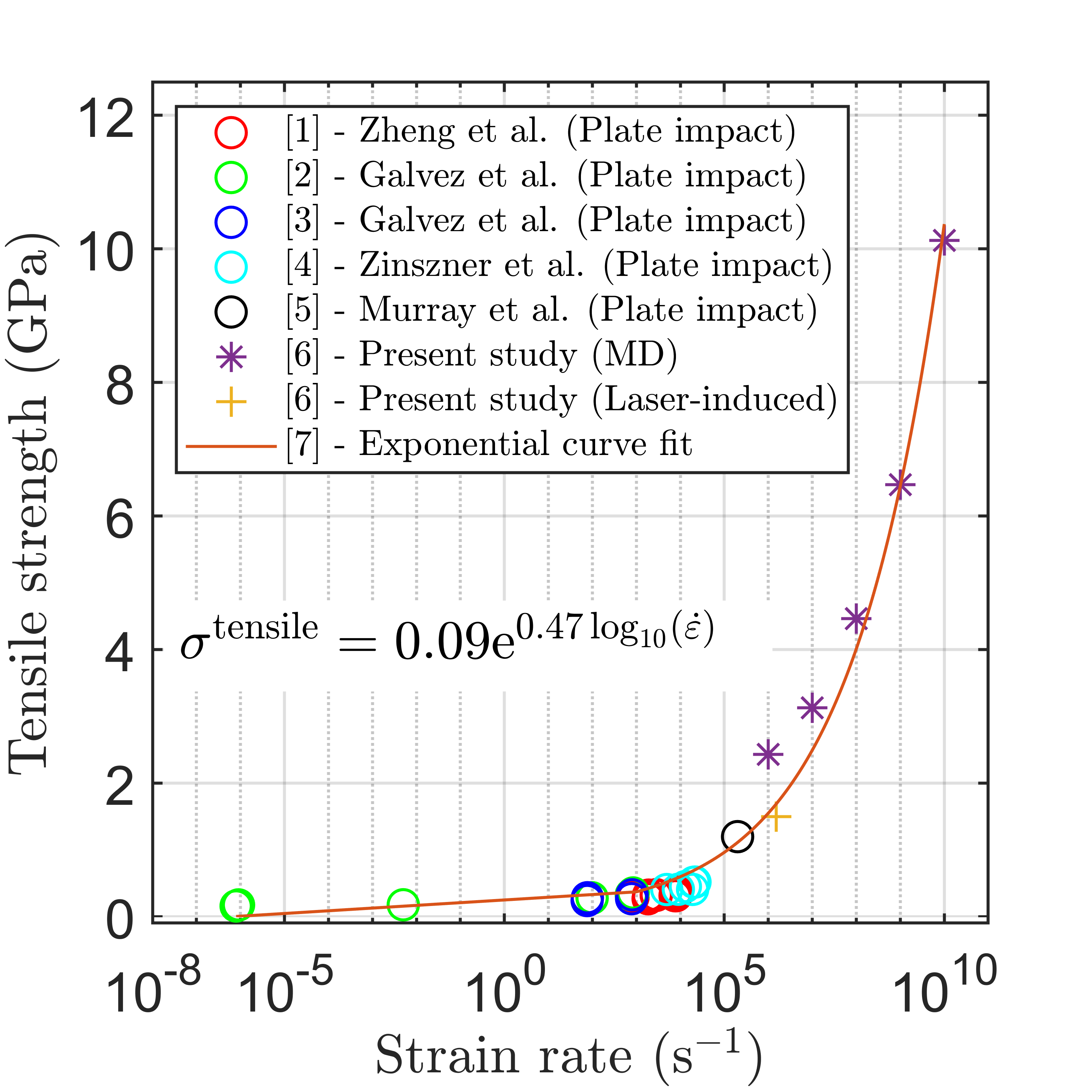}
\caption{Comparision of the spall strength for alumina for different strain rates. Data values combine plate impact experiments, the two-laser setup estimation and molecular dynamics simulations with the potential developed in this work. }
\label{fig:spall-strength}
\end{figure}
Figure \ref{fig:spall-strength} displays the tensile strength trend with strain rate for $\mathrm{Al_{2}O_{3}}$ across a range from $\mathrm{10^{-7}~s^{-1}}$ to $\mathrm{10^{10}~s^{-1}}$ with respect to the literature, MD simulations, and the laser-induced spall experiment.
The spall strength, i.e., essentially the tensile strength under high-strain-rates, exhibits good agreement with MD simulations and literature data, as evidenced by the exponential curve fit. 
Experimental studies in the literature have been limited to strain rates up to $\mathrm{10^{5}~s^{-1}}$ \cite{Zheng2023,Glvez2000,GlvezDazRubio2002,Zinszner2015,Murray1998}.
Under the single-pulse approach, prior studies have shown that spall strength can be inferred from the spall thickness, i.e., the distance of spall failure from the back surface \cite{gilath1988laser, Jarmakani2010, Mewael_1st_paper}.
Here, location 4 in Figure \ref{fig:Experiment_panel_2_b} corresponds to the largest void nearest to the back surface ($\sim800~\mu^3$, see Figure \ref{fig:pore_distribution_panel_single_two_pulse}), with a spall thickness of $\sim8\mu$m.
The spall thickness $\beta$ can be related to the shock velocity $V_s$, sample thickness $\omega$, bulk sound speed $c_0$, and pulse duration $\tau$ with the following relationship \cite{Mewael_1st_paper}:
\begin{equation} \label{eq:1}
V_s = \frac{c_0 \omega + \beta c_0}{\beta - \omega + \tau c_0}
\end{equation}
With $\omega = 250~\mu$m, $\tau = 350$ fs, $c_0$ = 7,455 $\mathrm{m \cdot s^{-1}}$, the calculated $V_s$ is 796 $\mathrm{m \cdot s^{-1}}$.
Using the Hugoniot relationship, the calculated particle velocity $V_p = (V_s - c_o)/\lambda$ = 387 $\mathrm{m \cdot s^{-1}}$ and shock pressure $P_o = \rho V_s V_p$ = 12 GPa. 
Here, the Huigonoit slope $\lambda$ is 1.31 \cite{Reinhart2003}.
For a triangular pressure pulse, the spall strength and strain rate can be expressed as $\sigma_\mathrm{spall} = 2(P_o - \rho c_o V_p)$ and $\dot{e} = U_p/L$, respectively \cite{gilath1988laser}.
From $\beta$, the spall strength in the $\mathrm{Al_{2}O_{3}}$ sample is estimated at $\sim1.5$ GPa, with an associated strain rate of $1.5\times 10^6 ~\mathrm{s}^{-1}$.
While the experimentally estimated spall strength is lower than the MD-predicted value ($\sigma_\text{spall}^\text{experiment}$ = 1.5 GPa and $\sigma_\text{spall}^\text{MD}$ = 2.4 GPa), this difference is expected, as MD simulations do not account for factors such as compression, impurities, and other complexities known to influence spall strength \cite{Louro1989, Murray1998,Bourne2001,Girlitsky2014,Hayun2018,Lebar2021}.

\section{Conclusions}
In this study, the spall behavior of $\mathrm{Al_{2}O_{3}}$ was examined using a combined approach that integrates high-fidelity informed molecular dynamics simulations with femtosecond laser-induced experiments.
An interatomic potential for $\mathrm{Al_{2}O_{3}}$ was developed by training an equivariant neural network model on an extensive dataset generated from density-functional theory calculations.
The molecular dynamics simulations explored a range of strain rates from $\mathrm{10^{6}~s^{-1}}$ to $\mathrm{10^{10}~s^{-1}}$, while the experiments were conducted at a strain rate on the order of $\mathrm{10^{6}~s^{-1}}$, with a maximum laser pulse energy of 90 $\mu$J.
To induce spall conditions, both single- (illumination of the front surface only) and two-pulse (illumination of both front and back surfaces) laser approaches were employed after coating a thin layer of nickel to the $\mathrm{Al_{2}O_{3}}$ samples to enhance energy absorption.
The main findings of this investigation are as follows:
\begin{itemize}
\item {The surface response of the Ni-coated $\mathrm{Al_{2}O_{3}}$ sample to single- and two-pulse laser approaches demonstrated that sufficient tensile stress induced Ni-$\mathrm{Al_{2}O_{3}}$ interface failure when the laser fluence surpassed a threshold. SEM analysis revealed a reduction in damage region size with decrease in laser fluence, correlating with decreased tensile stress (c.f., Figures \ref{fig:Experiment_panel_1_a} and \ref{fig:Experiment_panel_1_c}). The dynamic tensile stress primarily caused fracture at the interface rather than within the Ni film itself (see Figure \ref{fig:surface_observation_close_up_laser_fluence_1800_and_400_J_m2}), indicating that the local adhesion between Ni and $\mathrm{Al_{2}O_{3}}$ was weaker than the spall strength of Ni}.
\item {To investigate the impact of single- and two-pulse laser approaches on the spall behavior of bulk $\mathrm{Al_{2}O_{3}}$, micro-CT analysis was conducted on samples loaded with (a) single-pulse laser with a fluence of 1,800 $\mathrm{J\cdot m^{-2}}$ and (b) two-pulse approach with a reduced fluence of 600 $\mathrm{J\cdot m^{-2}}$ (see Figures \ref{fig:Experiment_panel_2_a} and \ref{fig:Experiment_panel_2_c}). The 3D volume reconstructions revealed that the two-pulse approach exhibited fewer voids predominantly, however, despite this, spallation conditions were still achieved. The single-pulse approach generated voids primarily close to the front and back surfaces, while the two-pulse approach produced voids at the sample's center (see Figures \ref{fig:Experiment_panel_2_b} and \ref{fig:Experiment_panel_2_d}). The estimated strain rate of $1.5\times 10^7 ~\mathrm{s}^{-1}$ and spall strength of 1.5 GPa under the single-pulse case were consistent with the molecular dynamics simulations and the literature \cite{Zheng2023,Glvez2000,GlvezDazRubio2002,Zinszner2015,Murray1998} (see Figure \ref{fig:Experiment_panel_2_f}).}
\item {MD simulations demonstrate that $\mathrm{Al_{2}O_{3}}$ exhibits strain rate-dependent spall behavior under hydrostatic loading, with spall strength and critical strain increasing as the strain rate rises, as shown in Figure \ref{fig:Molecular_dynamics_panel_b}. The exponential relationship between spall strength and strain rate, along with the increase in energy required for failure, points to the material's enhanced resistance due to suppressed relaxation processes and accumulated internal stresses resulting in a higher threshold for failure with increased strain rate.}
\item {MD simulations indicate that beyond the critical strain, bond breakage initiates near a nanovoid (see Figure \ref{fig:Molecular_dynamics_bond_breakage}), leading to brittle failure. The coordination number analysis indicates the transition from $\mathrm{AlO_{6}}$ to $\mathrm{AlO_{4}}$ and $\mathrm{AlO_{5}}$ with increasing strain, signifying bond cleavage (see Figure \ref{fig:Molecular_dynamics_panel_c}). This bond breakage forms micro-cracks that coalesce and propagate, forming larger cracks that eventually develops to form a spall plane (see Figures \ref{fig:Molecular_dynamics_panel_d}, and  \ref{fig:Molecular_dynamics_failure_XY_XZ_plane}---\ref{fig:Molecular_dynamics_failure_surf_mesh}). The failure mechanism is influenced by stress concentration, anisotropic behavior, and nanovoid size, with the literature validating these observations \cite{Louro1989,Bourne2001,Antoun2003}.}
\end{itemize}
Thus, our findings shed light on the spall behavior of $\mathrm{Al_{2}O_{3}}$ under high-strain-rates. 
The study highlights the unique advantages of femtosecond laser-induced spall approaches, particularly the two-pulse laser approach, which effectively reduces the laser fluence required to achieve sufficient spall conditions.
This study shows a comparison between the behavior of shocked specimens under single and two-pulse experiments with femtosecond laser-induced spall approaches.
Remarkably, we illustrated the ability of the two-pulse experiment to develop voids in the middle of the sample due to shock wave interactions. 
Our work opens up the possibility to further investigate the shock behavior of materials and interfaces that cannot be accessed through traditional experiments such as plate impact by controlling the spall failure location.  
Moreover, this approach enables the characterization of spall behavior at extremely high-strain-rates, surpassing the limitations of traditional plate-impact methods.
Coupling these experiments with high-fidelity molecular dynamics simulations allows for understanding the failure mechanisms that take place at \textbf{fs}--\textbf{ns} time and \textbf{nm} spatial scales.
Hence, the femtosecond laser-induced spall approach serves as a promising tool for analyzing the spall behavior of ceramics under conditions previously inaccessible with conventional techniques, paving the way for more efficient and high-throughput experimental methods.
\section*{Code availability}
The code developed to analytically calculate the thermoelastic stress and prepare the DFT simulations and data set can be found at:
\begin{itemize}
	\item \url{https://github.com/mewael-isiet/analytical-thermoelastic-wave}
	\item \url{https://github.com/MusannaGalib/AtomProNet}
\end{itemize}
\section*{CRediT author statement}
\textbf{Mewael Isiet}: Methodology, Software, Validation, Formal analysis, Investigation, Data Curation, Visualization, Writing - Original Draft, Writing - Review \& Editing. \textbf{Musanna Galib}: Methodology, Software, Validation, Data Curation, Writing - Original Draft, Writing - Review \& Editing. \textbf{Jerry I. Dadap}: Methodology, Supervision, Writing - Review \& Editing. \textbf{Yunhuan Xiao}: Methodology, Writing - Review \& Editing. \textbf{Ziliang Ye}: Resources, Writing - Review \& Editing. \textbf{Mauricio Ponga}: Conceptualization, Methodology, Software, Resources, Project administration, Funding acquisition, Supervision, Writing - Review \& Editing.
\section*{Data availability}
Data will be made available on request.
\section*{Acknowledgements}
We acknowledge the support of the New Frontiers in Research Fund (NFRFE-2019-01095) and the Natural Sciences and Engineering Research Council of Canada (NSERC) through the Discovery Grant and ALLRP 560447-2020 grants. This research was supported through high-performance computational resources and services provided by Advanced Research Computing at the University of British Columbia and the Digital Research Alliance of Canada. We acknowledge the support of Canada Foundation for Innovation (CFI). Z.Y. was supported by the Canada Research Chairs Program.
 \bibliographystyle{ieeetr} 
 \bibliography{main_ref}
\clearpage
\setcounter{section}{0}
\setcounter{equation}{0}
\setcounter{figure}{0}
\setcounter{table}{0}
\setcounter{page}{1}
\renewcommand{\thesection}{S\arabic{section}}
\renewcommand{\theequation}{S\arabic{equation}}
\renewcommand{\thefigure}{S\arabic{figure}}
\renewcommand{\thetable}{S\arabic{table}}
\renewcommand{\thepage}{SM\arabic{page}}
\renewcommand{\bibnumfmt}[1]{[S#1]}
\renewcommand{\citenumfont}[1]{S#1}
\begin{center}
	{\LARGE Supplemental Material \\Spall failure of alumina at high-strain rates using femtosecond laser experiments and high-fidelity molecular dynamics simulations}\\[10pt]
	{\large Mewael Isiet, Musanna Galib, Yunhuan Xiao, Jerry I. Dadap, Zliliang Ye, Mauricio Ponga} \\[5pt]
    \rule{\textwidth}{1pt} \\[5pt]
    \rule{\textwidth}{1pt} \\[5pt]
\end{center} 	
 \section*{Postmortem surface micrographs of interface failure}
 \begin{figure}[ht]
 	\centering
 	\begin{subfigure}[b]{0.32\textwidth}
 		\centering
 		\includegraphics[width=\linewidth]{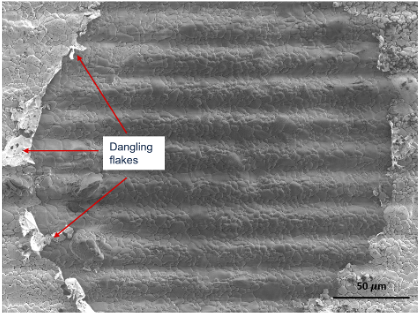}
 		\caption{Single pulse, 1,800 $\mathrm{J\cdot m^{-2}}$}
 	\end{subfigure}
 	\hfill
 	\begin{subfigure}[b]{0.32\textwidth}
 		\centering
 		\includegraphics[width=\linewidth]{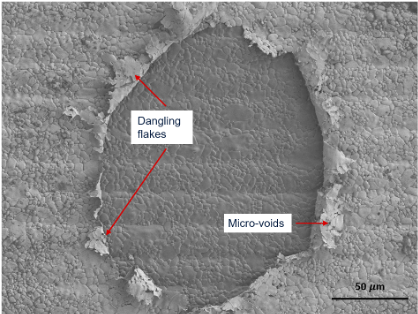}
 		\caption{Two pulse, 600 $\mathrm{J\cdot m^{-2}}$}
 	\end{subfigure}
 	\hfill
 	\begin{subfigure}[b]{0.32\textwidth}
 		\centering
 		\includegraphics[width=\linewidth]{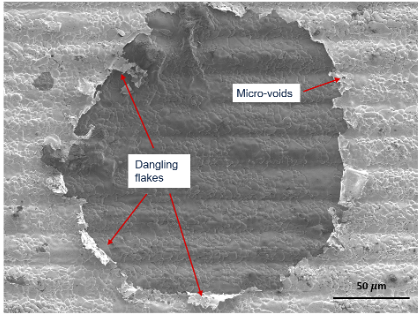}
 		\caption{Single pulse, 1,200 $\mathrm{J\cdot m^{-2}}$}
 	\end{subfigure}	
 	\hfill
 	\begin{subfigure}[b]{0.32\textwidth}
 		\centering
 		\includegraphics[width=\linewidth]{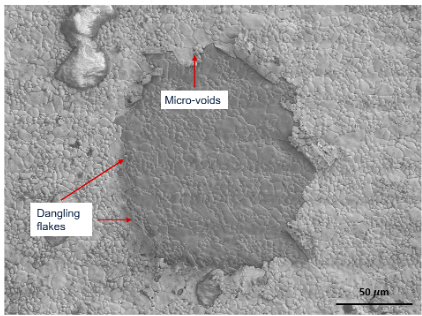}
 		\caption{Two pulse, 500 $\mathrm{J\cdot m^{-2}}$}
 	\end{subfigure}	
 	\hfill
 	\begin{subfigure}[b]{0.32\textwidth}
 		\centering
 		\includegraphics[width=\linewidth]{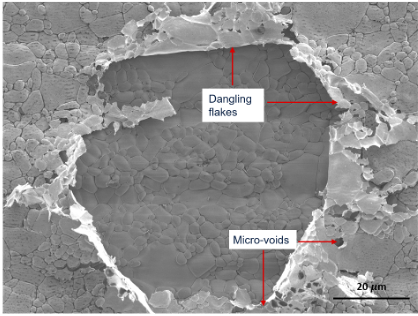}
 		\caption{Single pulse, 480 $\mathrm{J\cdot m^{-2}}$}
 	\end{subfigure}	
 	\hfill
 	\begin{subfigure}[b]{0.32\textwidth}
 		\centering
 		\includegraphics[width=\linewidth]{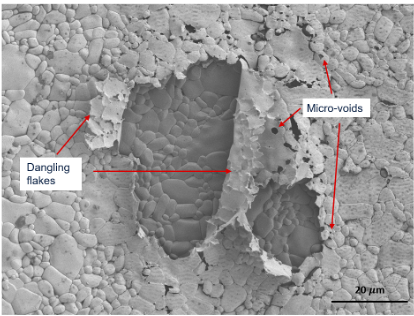}
 		\caption{Two pulse, 400 $\mathrm{J\cdot m^{-2}}$}
 	\end{subfigure}	
 	\caption{Closeup SEM images of the (front) surface features produced after single- and two-pulse laser illumination on the nickel coated $\mathrm{Al_{2}O_{3}}$ for a range of laser fluence between 1,800 to 400 $\mathrm{J\cdot m^{-2}}$. In these images, the Ni coating (visible as silver toned) is displayed with the underlying $\mathrm{Al_{2}O_{3}}$ substrate exposed with varying degrees depending on the laser fluence. Furthermore, dangling Ni flakes and micro-voids on the Ni coating can be seen under both approaches.}
 	\label{fig:surface_observation_close_up_laser_fluence_1800_and_400_J_m2}
 \end{figure}
 \newpage
 \section*{Pore distribution in post laser-shocked alumina}
 \begin{figure}[ht]
 	\centering
 	\begin{subfigure}[b]{0.2\textwidth}
 		\textbf{Single pulse}\\[0.5em]
 		\centering
 		\includegraphics[width=\linewidth]{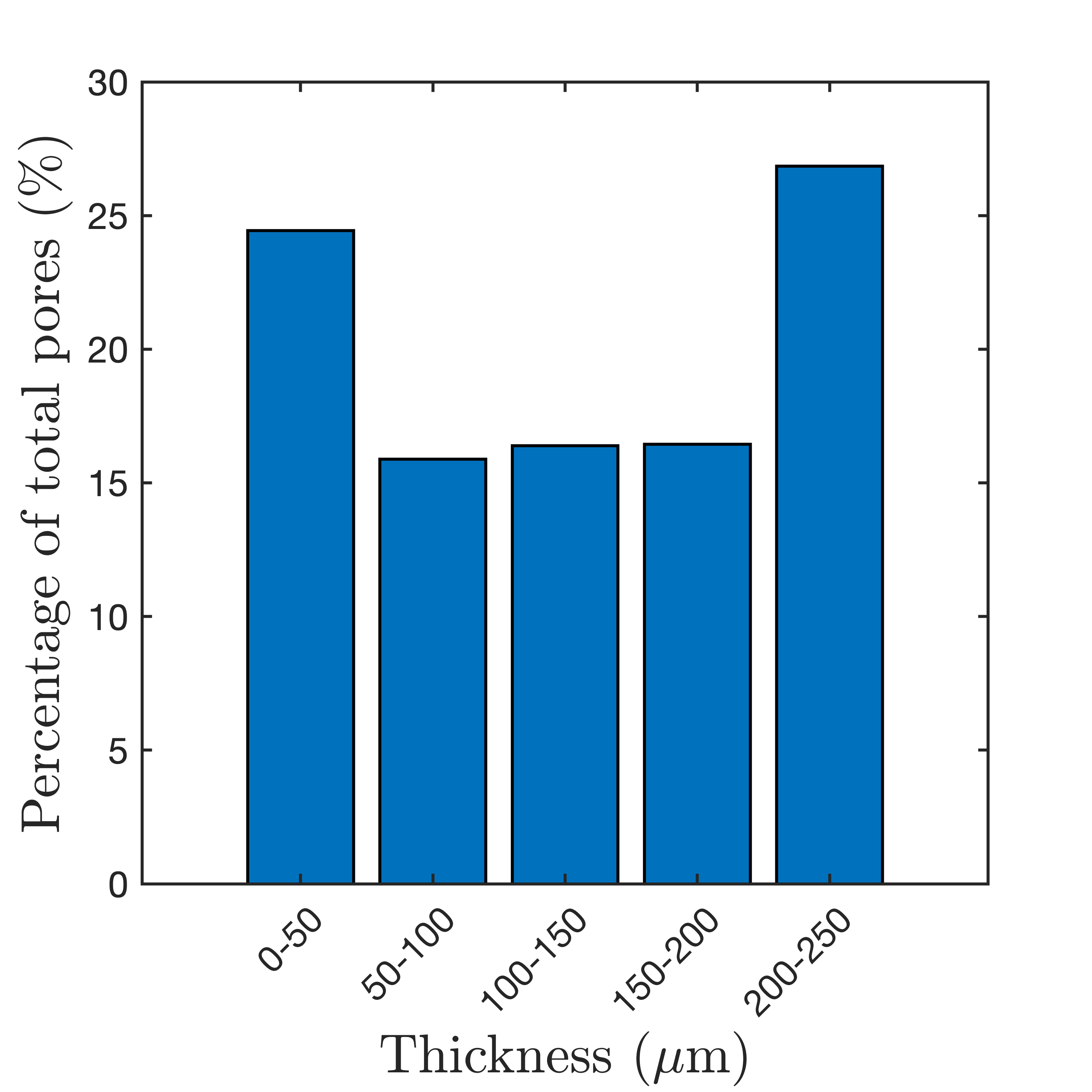}
 		\caption{1,800 $\mathrm{J\cdot m^{-2}}$}
 		\label{fig:pore_distribution_1800_J_m2_single_pulse}
 	\end{subfigure}
 	\hfill
 	\begin{subfigure}[b]{0.2\textwidth}
 		\textbf{Single pulse}\\[0.5em]    
 		\centering
 		\includegraphics[width=\linewidth]{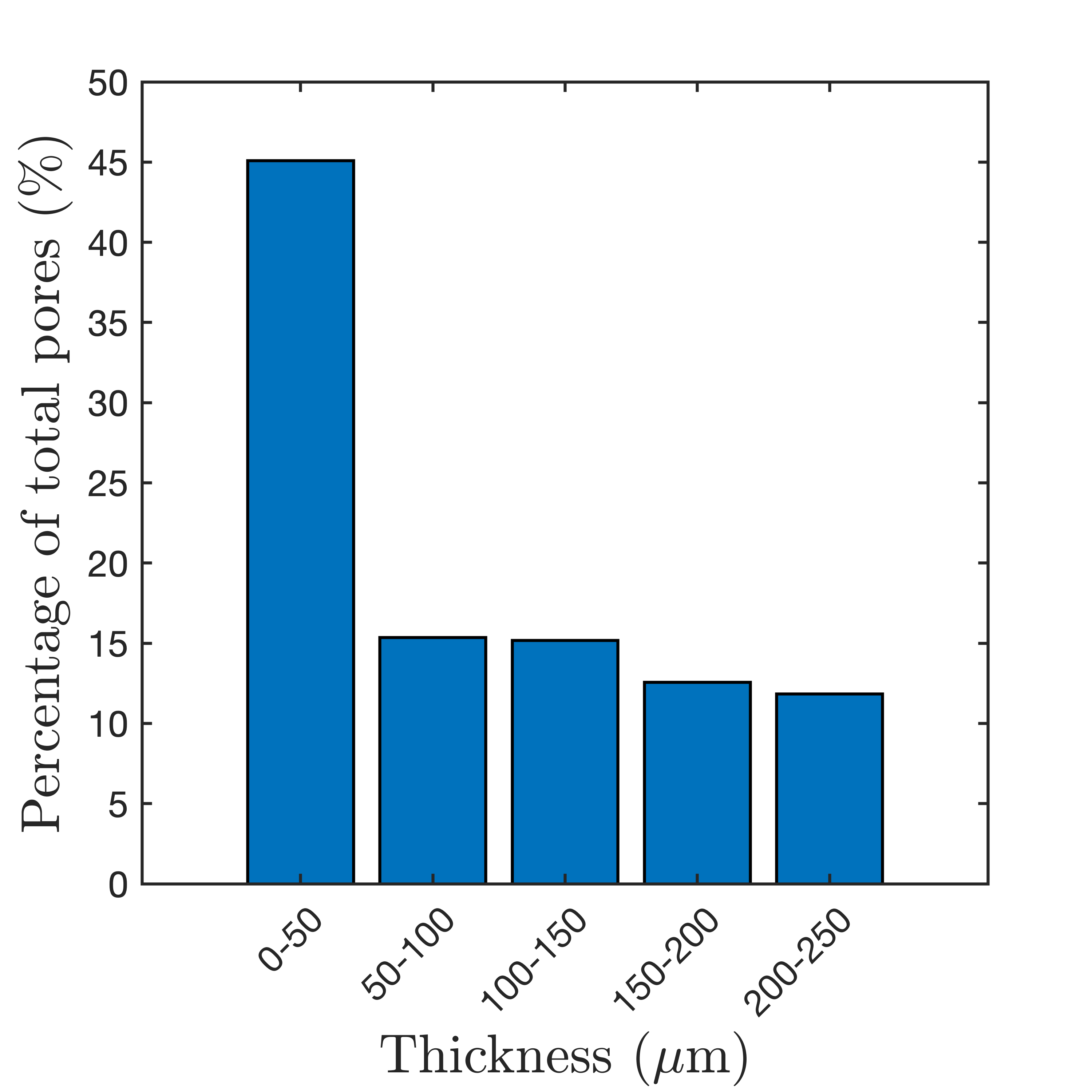}
 		\caption{1,200 $\mathrm{J\cdot m^{-2}}$}
 		\label{fig:pore_distribution_1200_J_m2_single_pulse}
 	\end{subfigure}
 	\hfill    
 	\begin{subfigure}[b]{0.2\textwidth}
 		\textbf{Two pulse}\\[0.5em]    
 		\centering
 		\includegraphics[width=\linewidth]{Figures/pore_distribution_600_J_m2_two_pulse.png}
 		\caption{600 $\mathrm{J\cdot m^{-2}}$}
 		\label{fig:pore_distribution_600_J_m2_two_pulse}
 	\end{subfigure}
 	\hfill
 	\begin{subfigure}[b]{0.2\textwidth}
 		\textbf{Two pulse}\\[0.5em]    
 		\centering
 		\includegraphics[width=\linewidth]{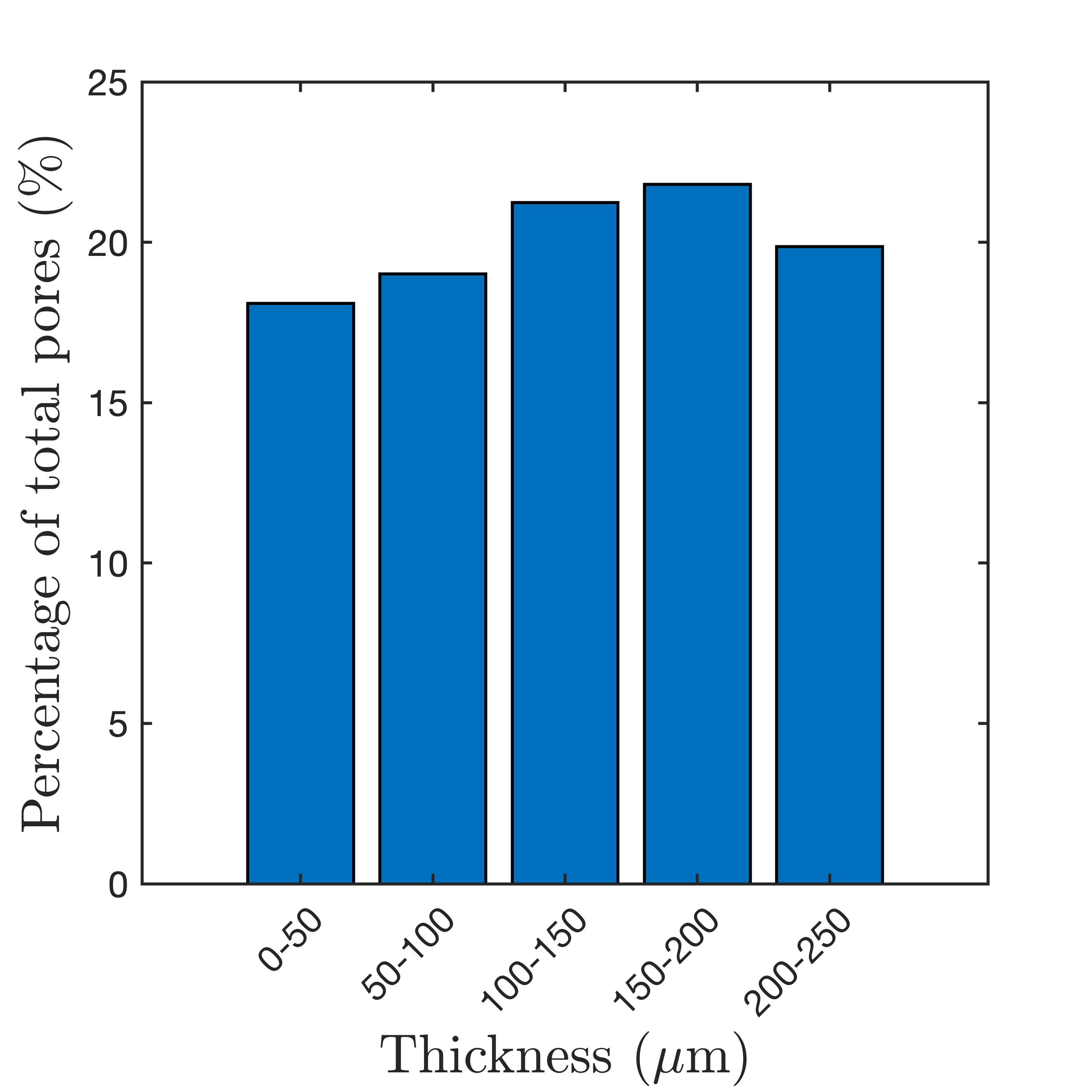}
 		\caption{500 $\mathrm{J\cdot m^{-2}}$}
 		\label{fig:pore_distribution_500_J_m2_two_pulse}
 	\end{subfigure}
 	\hfill    
 	\begin{subfigure}[b]{0.2\textwidth}
 		\centering
 		\includegraphics[width=\linewidth]{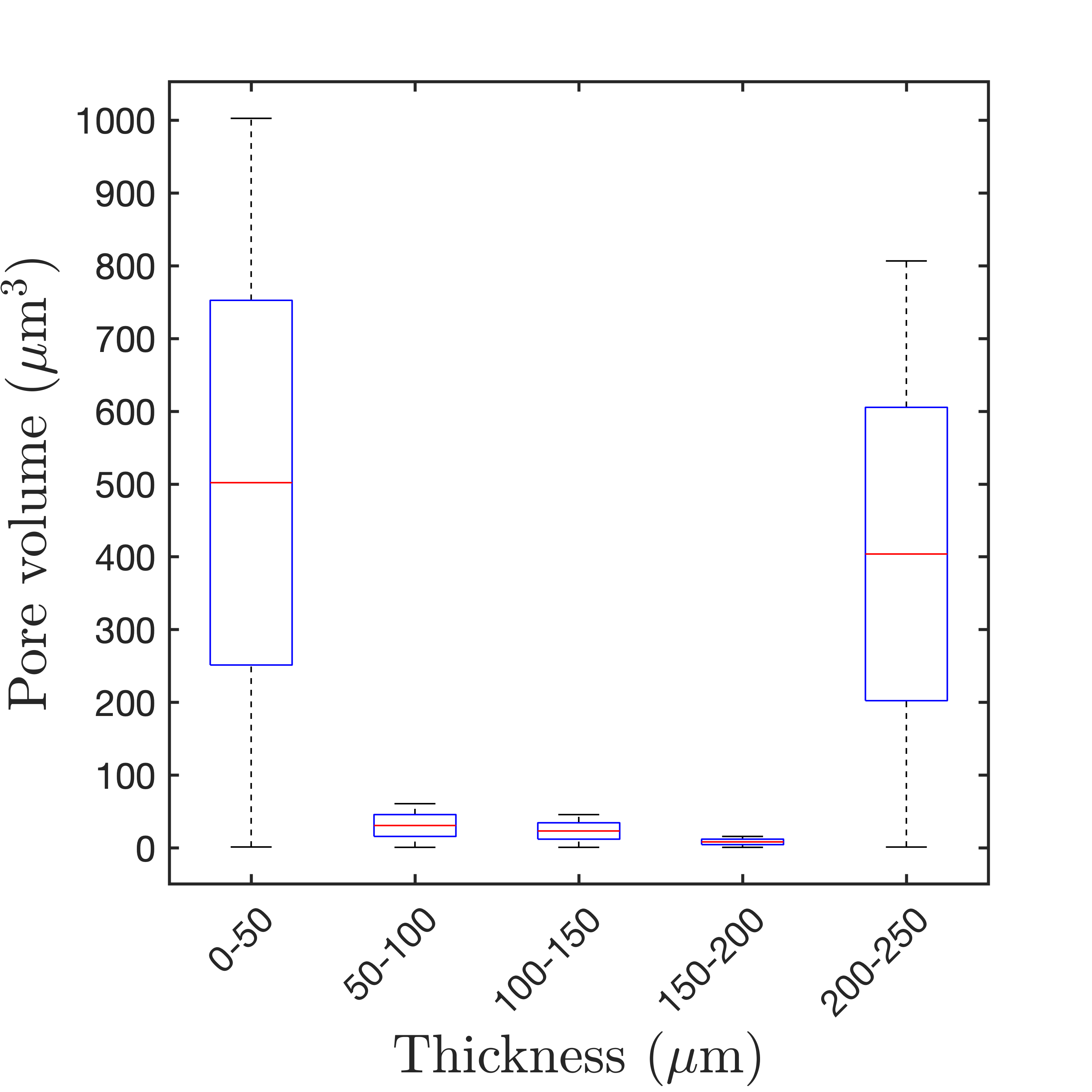}
 		\caption{1,800 $\mathrm{J\cdot m^{-2}}$}
 		\label{fig:pore_volume_distribution_1800_J_m2_single_pulse}
 	\end{subfigure}
 	\hfill
 	\begin{subfigure}[b]{0.2\textwidth}
 		\centering
 		\includegraphics[width=\linewidth]{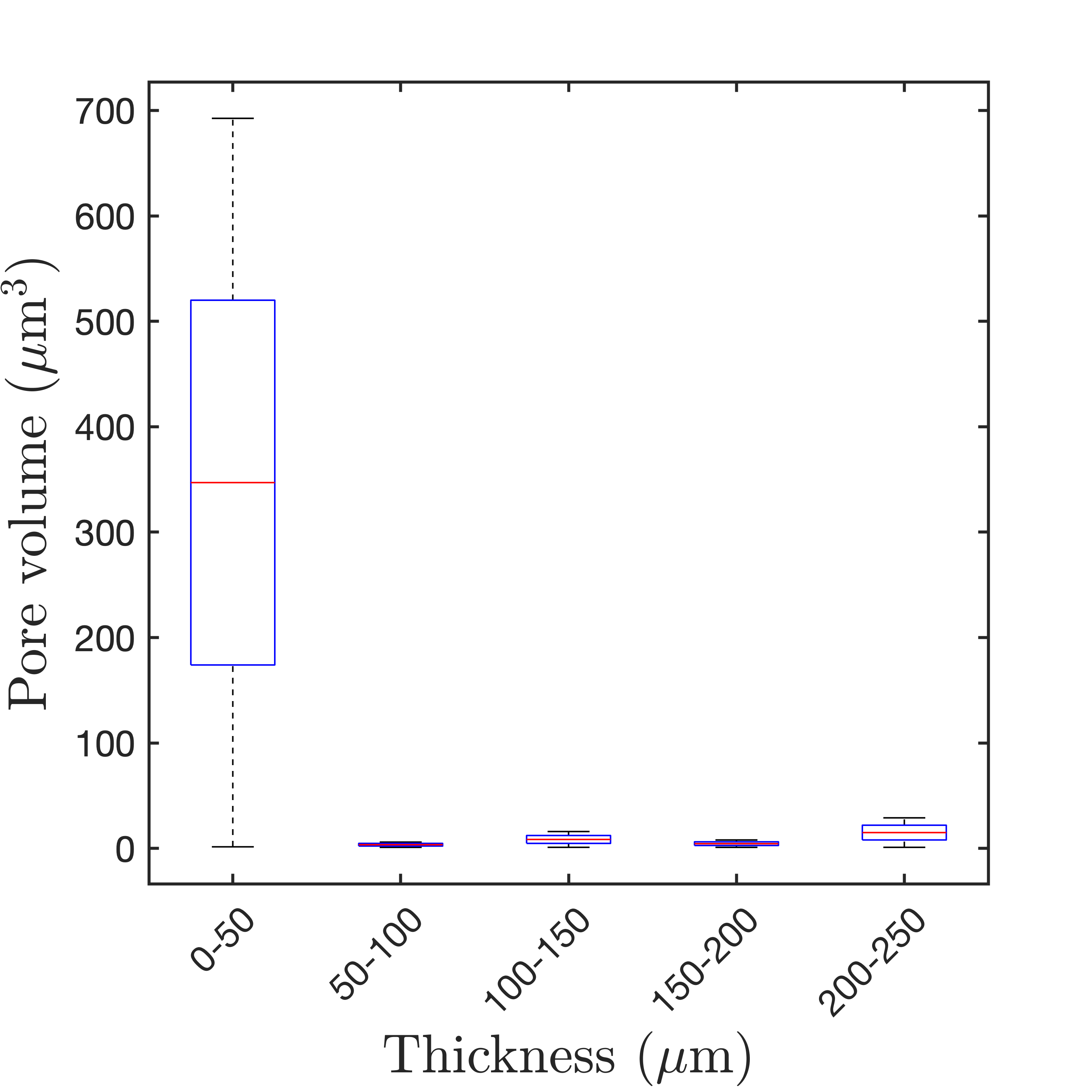}
 		\caption{1,200 $\mathrm{J\cdot m^{-2}}$}
 		\label{fig:pore_volume_distribution_1200_J_m2_single_pulse}
 	\end{subfigure}
 	\hfill    
 	\begin{subfigure}[b]{0.2\textwidth}
 		\centering
 		\includegraphics[width=\linewidth]{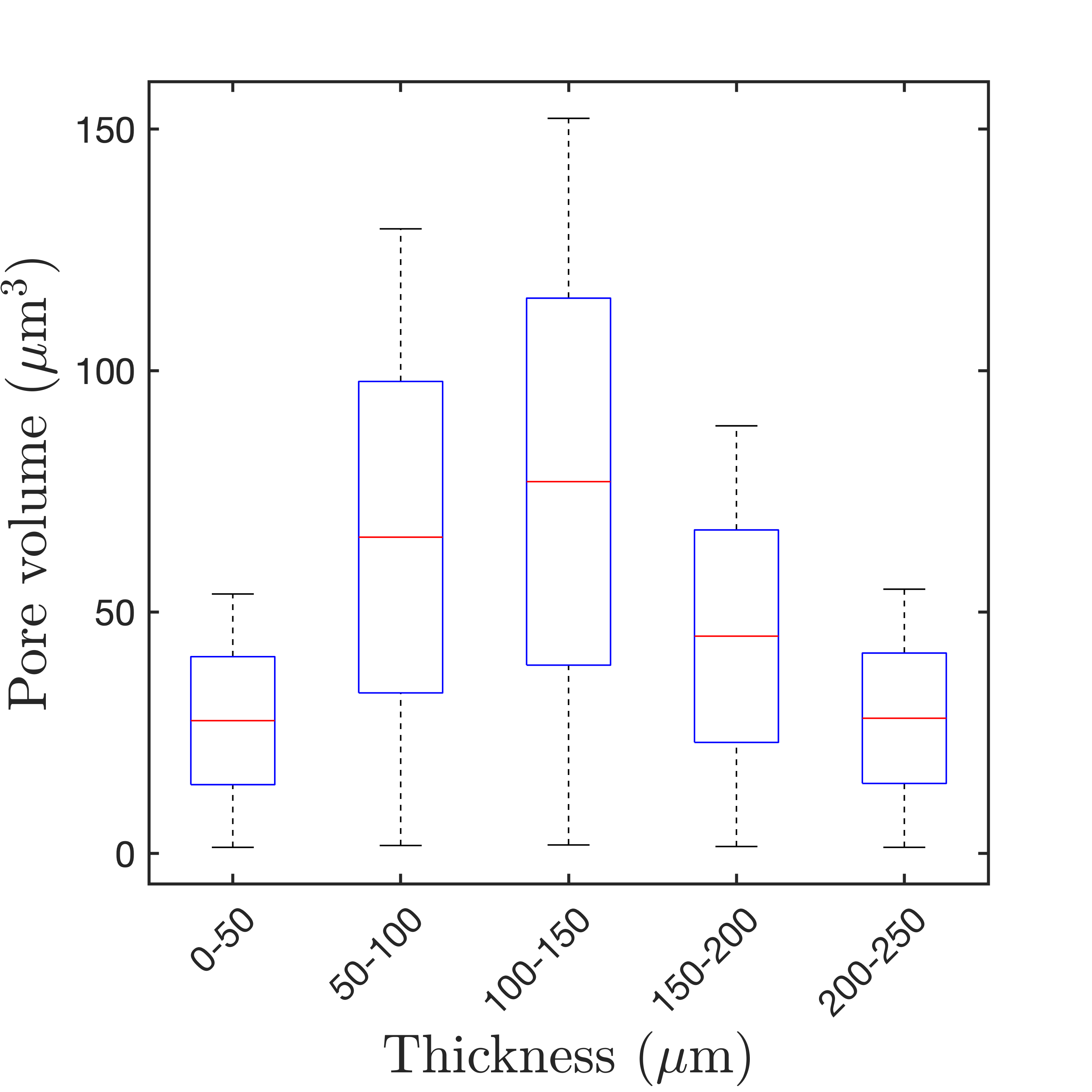}
 		\caption{600 $\mathrm{J\cdot m^{-2}}$}
 		\label{fig:pore_volume_distribution_600_J_m2_two_pulse}
 	\end{subfigure}
 	\hfill    
 	\begin{subfigure}[b]{0.2\textwidth}
 		\centering
 		\includegraphics[width=\linewidth]{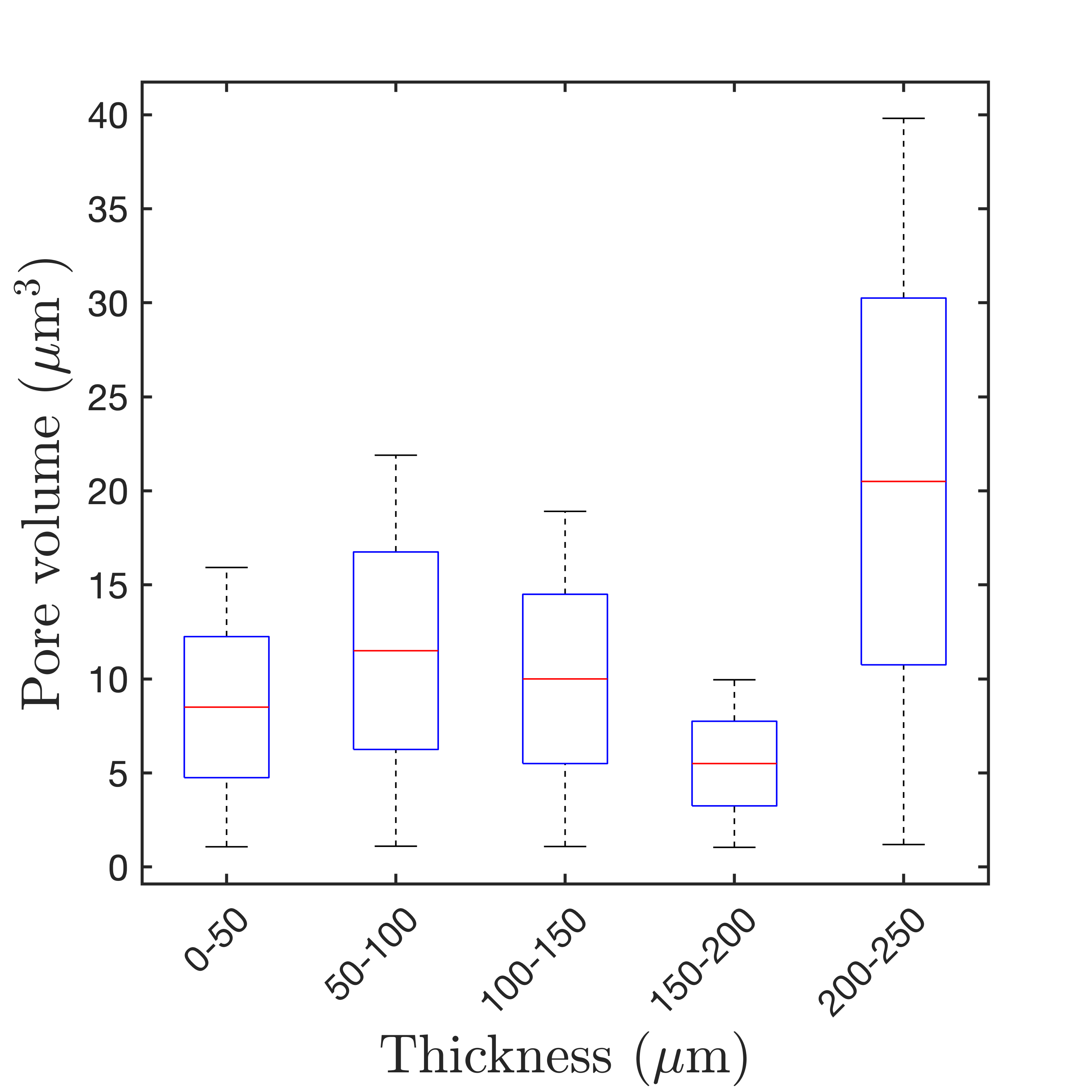}
 		\caption{500 $\mathrm{J\cdot m^{-2}}$}
 		\label{fig:pore_volume_distribution_500_J_m2_two_pulse}
 	\end{subfigure}        
 	\caption{Porosity analysis of the $\mathrm{Al_{2}O_{3}}$ sample illuminated by the single- and two-pulse laser approaches with (a--d) highlighting the distribution of pore count percentage as a function of thickness and (e--h) illustrating the boxplot of pore volume distribution across thickness, highlighting statistical variations. These results provide insight into the two-pulse approach ability to nucleate voids at the centre of the sample similar to the plate-impact method.}
 	\label{fig:pore_distribution_panel_single_two_pulse}
 \end{figure}
 The effect of the single- (fluence: 1,800 and 1,200 $\mathrm{J\cdot m^{-2}}$) and two-pulse approach (fluence: 600 and 500 $\mathrm{J\cdot m^{-2}}$) on the pore (void) distribution in $\mathrm{Al_{2}O_{3}}$ can be seen in Figure \ref{fig:pore_distribution_panel_single_two_pulse}.
 To investigate the pore distribution across thickness, multiple 50 nm-wide slices of the 3D reconstructed sample were created along the laser-pulse direction. 
 In the single-pulse approach at 1,800 $\mathrm{J\cdot m^{-2}}$, failure primarily occurred near the front and back surfaces (i.e., the 0-50 and 200-250 nm bins) due to the passage of the unloading tensile wave and the interaction of the reflected compressive wave with the unloading wave near the back surface (see Figures \ref{fig:pore_distribution_1800_J_m2_single_pulse}--\ref{fig:pore_volume_distribution_1800_J_m2_single_pulse} and \ref{fig:Experiment_panel_2_b}). 
 At 1,200 $\mathrm{J\cdot m^{-2}}$, failure was observed only near the front surface--closer to the illuminated surface--indicating that classical spall failure (under the single-pulse approach),which typically results from the interaction of the reflected compressive wave with the unloading wave near the back surface, did not occur in this case.
 As for the two-pulse approach under 600 $\mathrm{J\cdot m^{-2}}$, the interaction of the unloading tensile waves at the sample center generated large hydrostatic stress, leading to failure predominantly taking place at the center (corresponding to the 100-150 nm bin) as shown in Figure \ref{fig:pore_distribution_600_J_m2_two_pulse}.
 Additionally, bins closer to the center (50–100 nm and 150–200 nm) exhibited a higher void density than those near the surface.
 Furthermore, the median, largest pore volume, and interquartile range of the pore distribution followed a similar trend, as shown in Figure \ref{fig:pore_volume_distribution_600_J_m2_two_pulse}.
 At 500 $\mathrm{J\cdot m^{-2}}$, despite an increase in the median and largest pore volume at the back surface (200–250 nm) compared to other regions, minimal spall features were observed when comparing the porosity analysis with the pristine sample (see Figures \ref{fig:pristine_histogram} and \ref{fig:Experiment_panel_2_e}).
 Hence, this analysis highlights (a) the laser fluence threshold for spallation in the single- and two-pulse laser approaches, identified as 1,800 and 600 $\mathrm{J\cdot m^{-2}}$, respectively and (b) the ability of the two-pulse approach to produce central voids, which correlates to the observed failure patterns seen in the 2D micro-CT slices, as shown in Figure \ref{fig:Experiment_panel_2}.
 \newpage
 \section*{Interatomic potential stress-strain prediction for alumina}\label{inter_atomic_prediction}
 \begin{figure}[ht]
 	\centering
 	\begin{subfigure}[b]{0.45\textwidth}
 		\centering
 		\includegraphics[width=\linewidth]{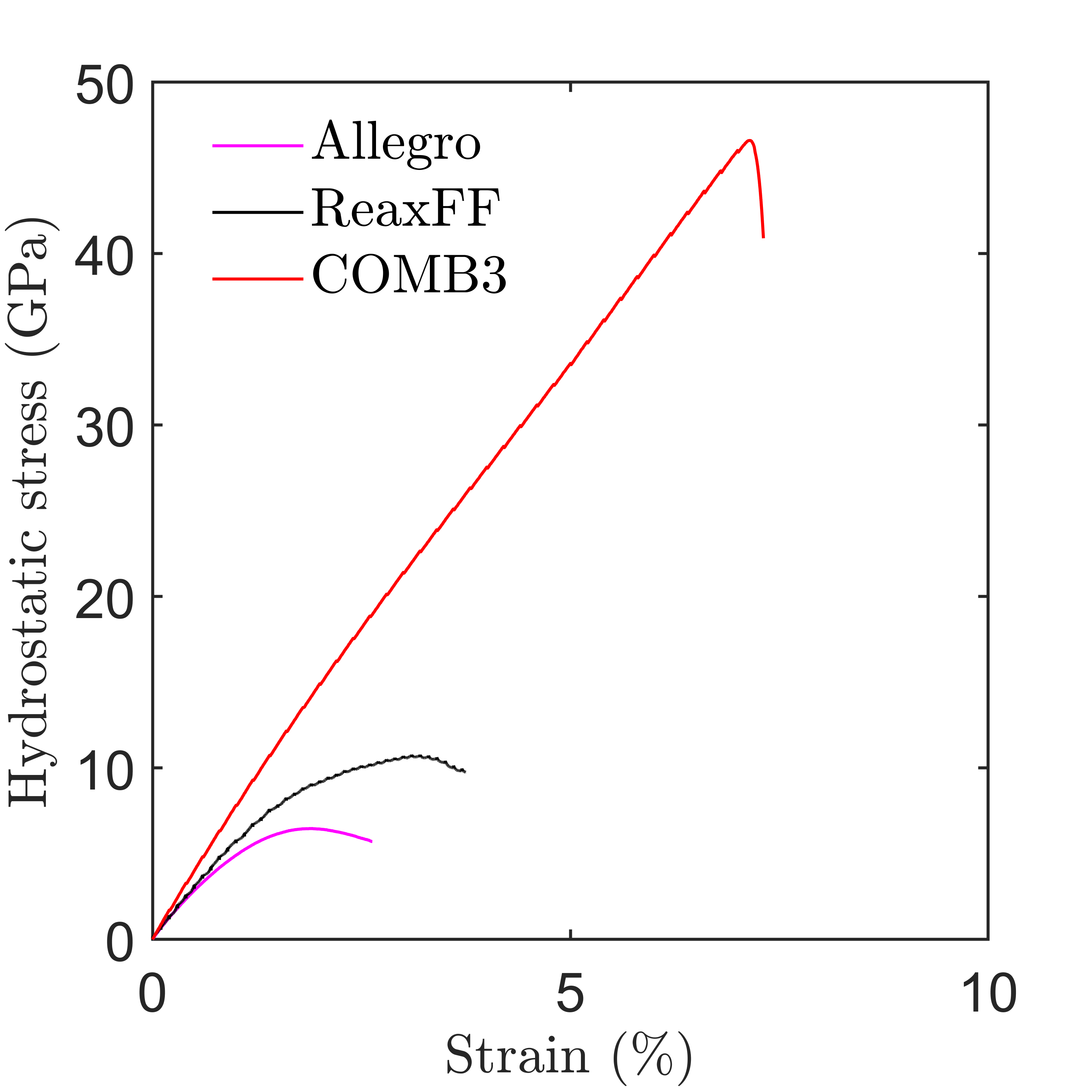}
 		\caption{}
 		\label{fig:Molecular_dynamics_COMB_ALLEGRO_REAXFF_comparison_a}
 	\end{subfigure}
 	\hfill
 	\begin{subfigure}[b]{0.45\textwidth}
 		\centering
 		\includegraphics[width=\linewidth]{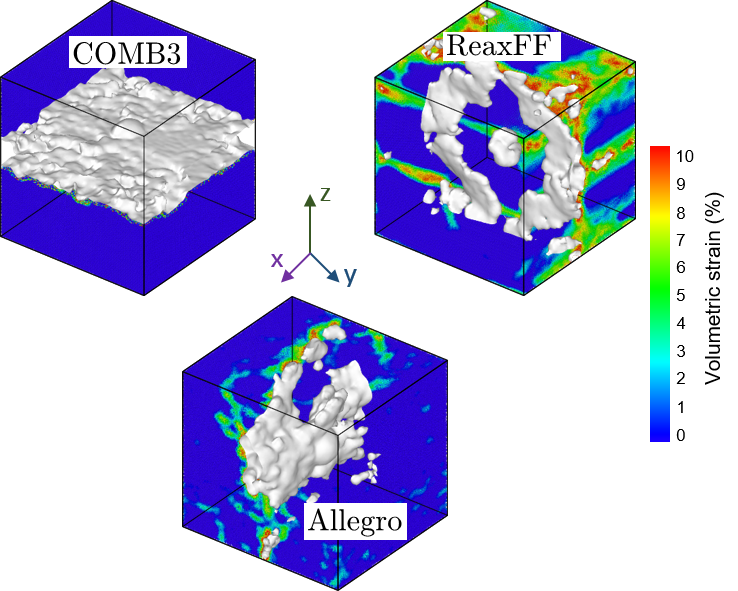}
 		\caption{}
 		\label{fig:Molecular_dynamics_COMB_ALLEGRO_REAXFF_comparison_b}
 	\end{subfigure}
 	\caption{Application of reactive, COMB3 and ReaxFF, and machine learning, Allegro, force fields to predict the (a) stress-strain behaviour and (b) failure pattern, at a strain of 5$\%$, of $\mathrm{Al_{2}O_{3}}$ under a strain rate of $\mathrm{10^{9}~s^{-1}}$ for a simulation cell with a volume of 30 $\mathrm{nm^{3}}$ and a nano-void diameter of 6 nm. A surface mesh, based on the alpha-shape method \cite{stukowski2012automated}, is imposed on the simulation cell to highlighted the failure pattern predicted under the molecular dynamics simulations.}
 	\label{fig:Molecular_dynamics_COMB_ALLEGRO_REAXFF_comparison}
 \end{figure}
 Figure \ref{fig:Molecular_dynamics_COMB_ALLEGRO_REAXFF_comparison_a} highlights the stress-strain curve calculated using three different interatomic potentials.
 It can be seen that the Allegro potential predicted a much lower peak stress, i.e., spall strength, in comparison to the other potentials.
 In comparision, the spall strength (and critical strain) predicted by Allegro is 6.5 GPa (1.9$\%$), whereas ReaxFF and COMB3 are 10.8 (3.2$\%$) and 46.8 GPa (7.1$\%$), respectively.
 It can be seen that all potentials predicted an initial linear trend, however it can be seen that slope of the stress-strain curves differ.
 On the one hand, the COMB potential predicts an almost constant slope, whereas the Allegro and ReaxFF predicted a reduction in the slope with increasing strain - this difference can be attributed to the elastic constant tensor dependence on deformation.

 The predicted failure pattern by the three force fields is given in Figure \ref{fig:Molecular_dynamics_COMB_ALLEGRO_REAXFF_comparison_b}.
 The COMB potential predicted a single failure plane along the XY plane, however, the Allegro and ReaxFF potentials predicted failure along the XZ and YZ plane, respectively.
 Also, the COMB and Allegro potentials correctly predicted failure at the vicinity of the nanovoid, where as the ReaxFF potential predicted failure away from the nanovoid.
 Lastly, bond breakage, prior to microcrack formation and propagation, was predicted by the Allegro and ReaxFF potentials, however, the COMB potential predicted nearly instantaneous failure once the critical strain was achieved.
 Overall, the Allegro potential managed to predict a lower spall strength and critical strain, while predicting a brittle failure involving bond breakage, microcrack formation, and crack propagation and showing good agreement with DFT calculated elastic constants in Table \ref{tb:Elastic_constant_comparison}.
 Hence, this makes the Allegro potential a good alternative to the other classical reactive force fields, i.e., ReaxFF and COMB.
 \newpage
 \section*{Bond breakage mechanism in alumina}
 \begin{figure}[ht]
 	\centering
 	\includegraphics[width=1\linewidth]{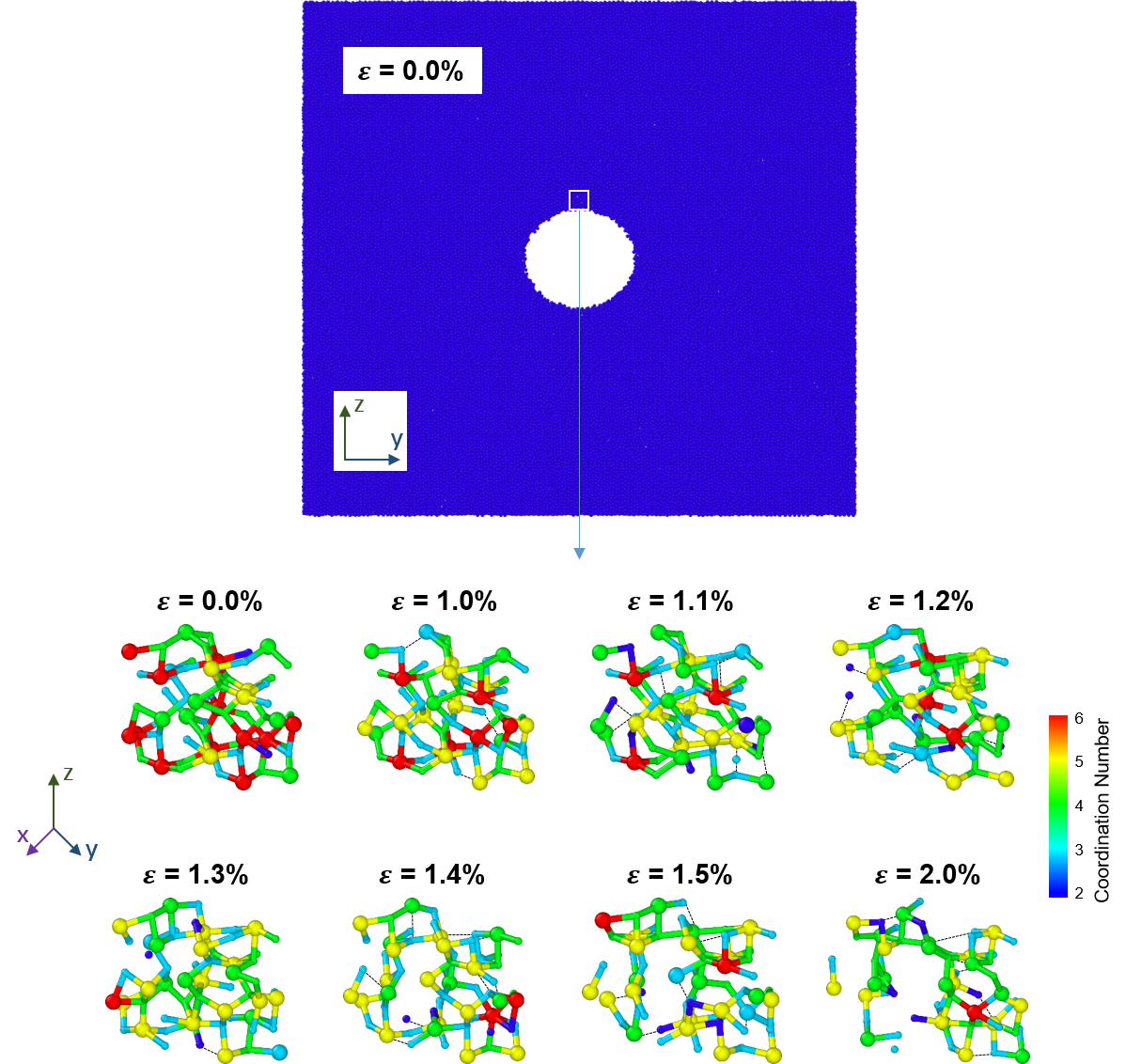}
 	\caption{Bond breakage in $\mathrm{Al_{2}O_{3}}$ predicted under hydro-static loading with a strain rate of $\mathrm{10^{8}~s^{-1}}$ for a simulation cell with a volume of 30 $\mathrm{nm^{3}}$ and a nano-void diameter of 6 nm. Here the black dashed line corresponds to a broken bond.}
 	\label{fig:Molecular_dynamics_bond_breakage}
 \end{figure}
 \newpage
 \section*{Failure behavior of alumina along XY and XZ plane}
 \begin{figure}[ht]
 	\centering
 	\includegraphics[width=1\linewidth]{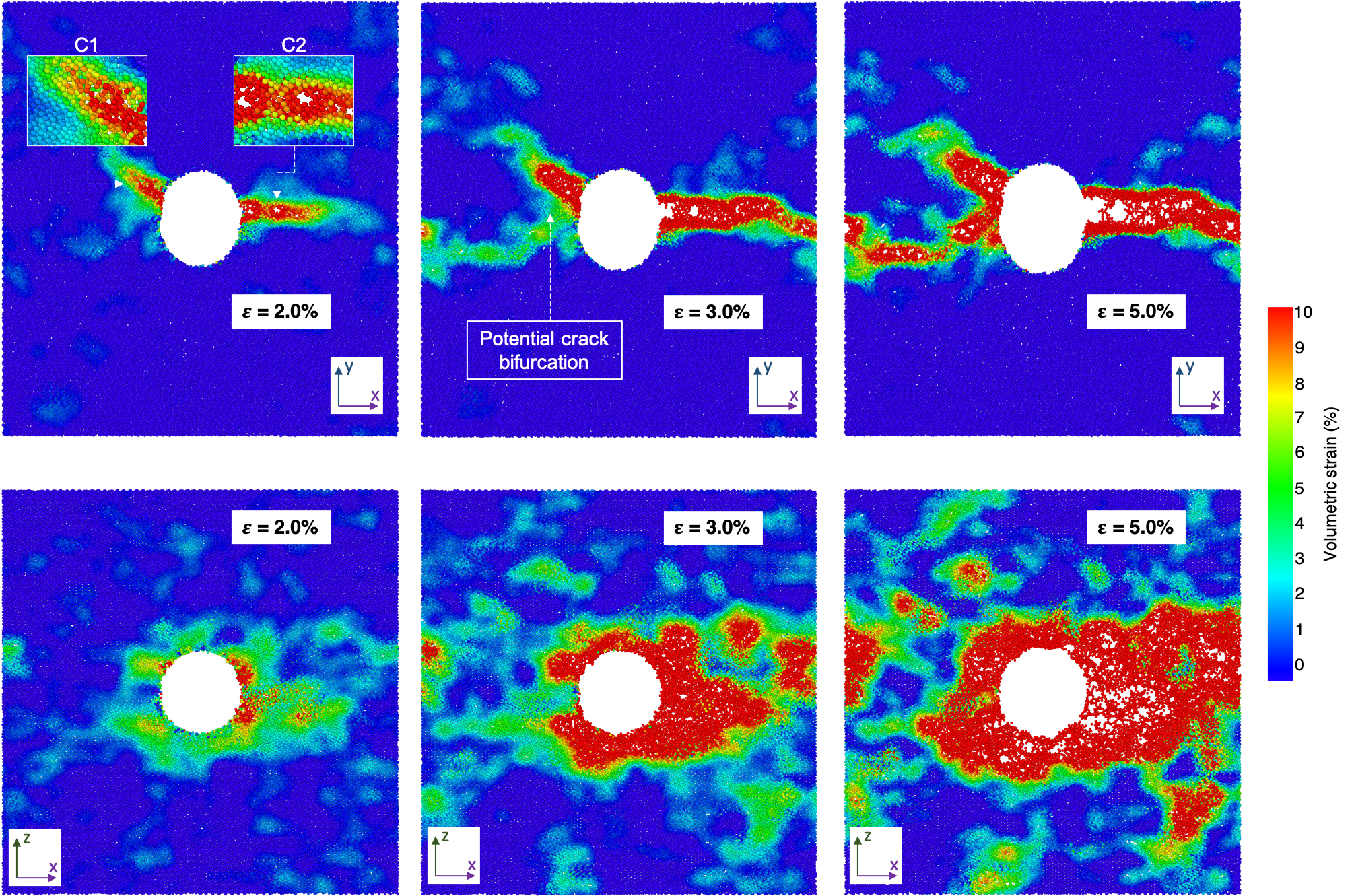}
 	\caption{Molecular dynamics simulation of $\mathrm{Al_{2}O_{3}}$ under hydrostatic loading highlighting the spall behaviour along the \textbf{XY} and \textbf{XZ} plane under a strain rate of $\mathrm{10^{8}~s^{-1}}$ for a simulation cell with a volume of 30 $\mathrm{nm^{3}}$ and a nano-void diameter of 6 nm.}
 	\label{fig:Molecular_dynamics_failure_XY_XZ_plane}
 \end{figure}
 \newpage
 \section*{Evolution of failure in alumina with increasing strain}
 \begin{figure}[ht]
 	\centering
 	\includegraphics[width=1\linewidth]{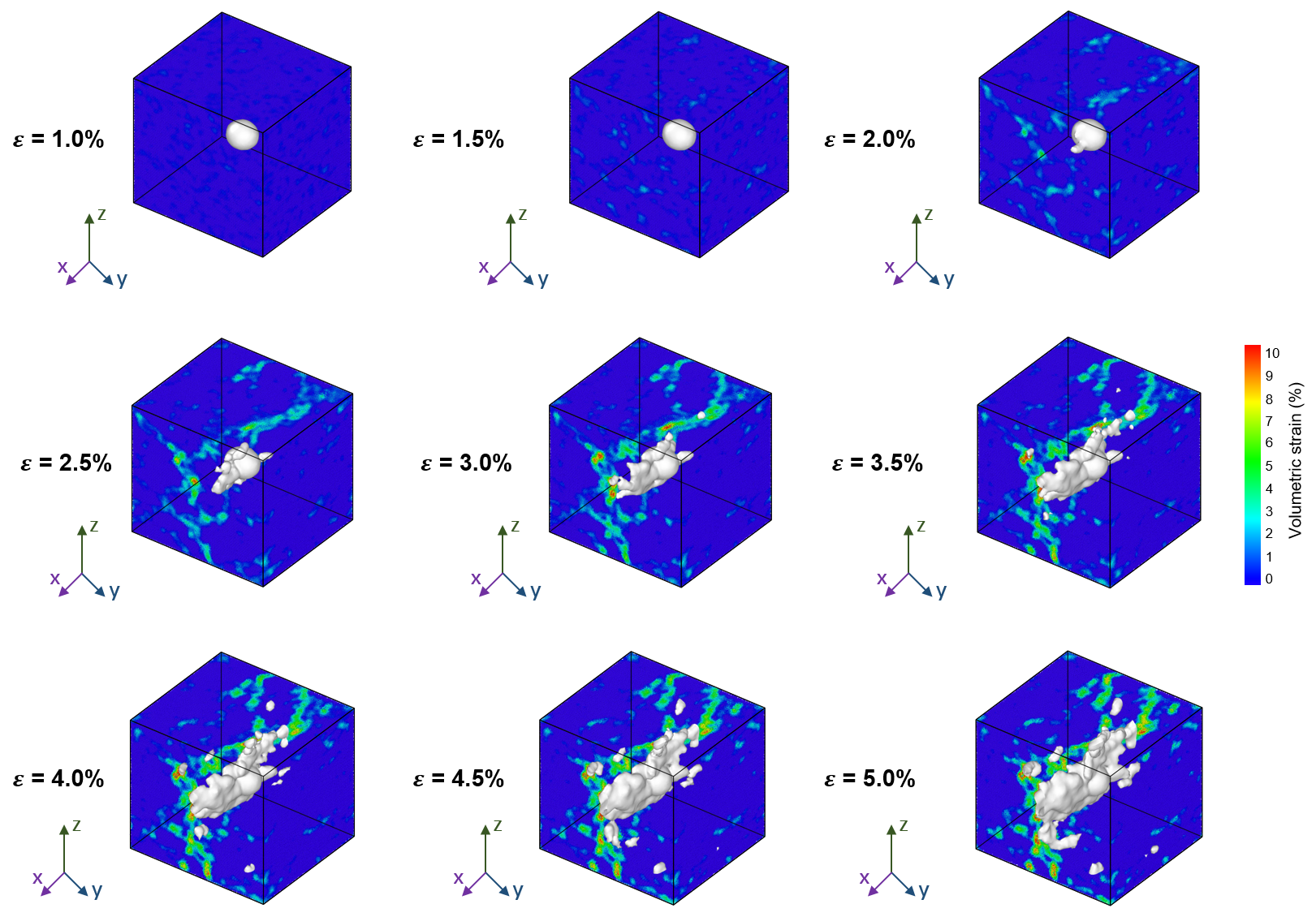}
 	\caption{Schematic of the brittle failure of $\mathrm{Al_{2}O_{3}}$ under hydrostatic loading with a strain rate of $\mathrm{10^{8}~s^{-1}}$ from 1.0 $\%$ to 5.0 $\%$ with a 6 nm diameter void. A surface mesh, based on the alpha-shape method \cite{stukowski2012automated}, is imposed on the simulation cell to highlighted the failure pattern predicted under the molecular dynamics simulations.}
 	\label{fig:Molecular_dynamics_failure_surf_mesh}
 \end{figure}
 \newpage
 \section*{Nanovoid size effect on spall behavior of alumina}
 \begin{figure}[h!]
 	\centering
 	\begin{subfigure}[b]{0.45\textwidth}
 		\centering
 		\includegraphics[width=\linewidth]{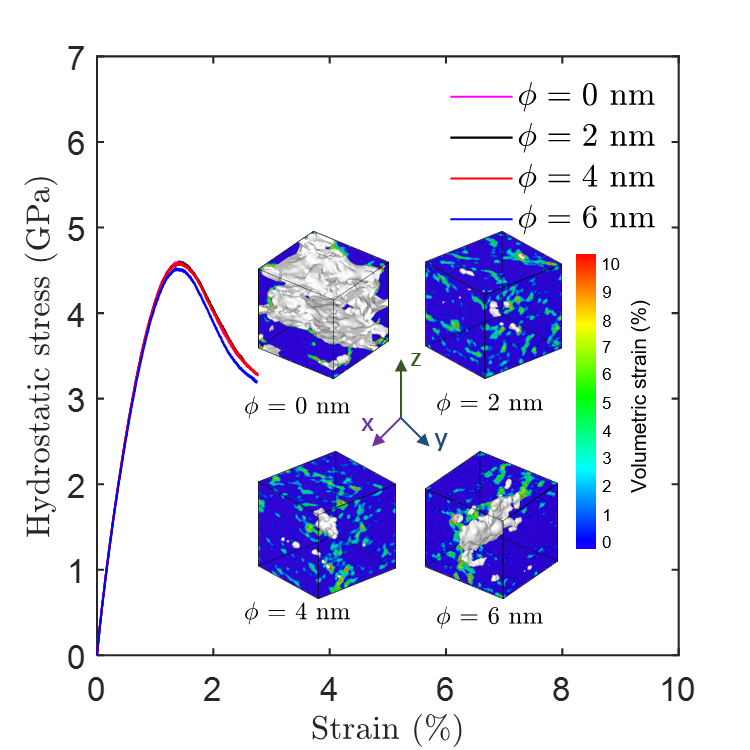}
 		\caption{}
 		\label{fig:Molecular_dynamics_void_size_strain_rate_a}
 	\end{subfigure}
 	\hfill
 	\begin{subfigure}[b]{0.45\textwidth}
 		\centering
 		\includegraphics[width=\linewidth]{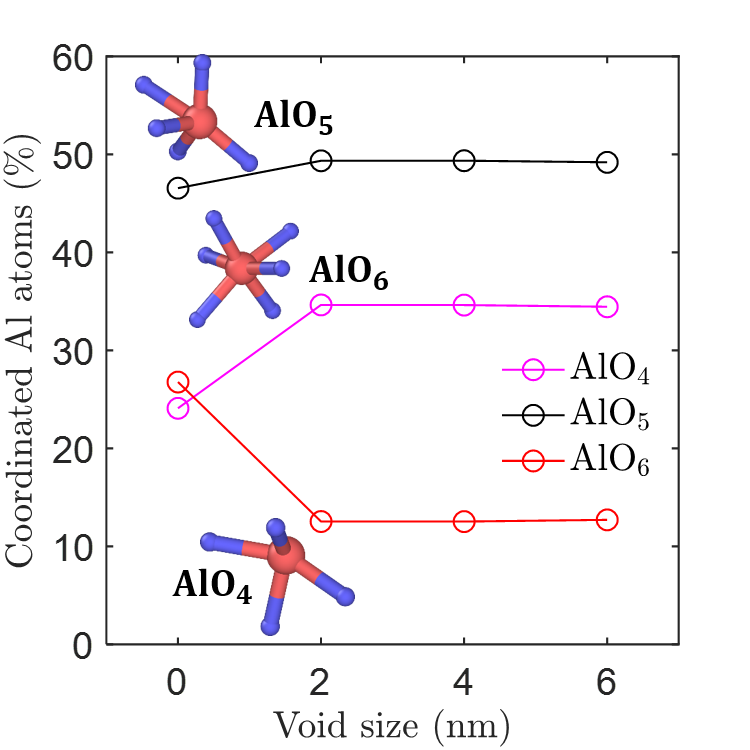}
 		\caption{}
 		\label{fig:Molecular_dynamics_void_size_strain_rate_b}
 	\end{subfigure}
 	\hfill
 	\begin{subfigure}[b]{0.45\textwidth}
 		\centering
 		\includegraphics[width=\linewidth]{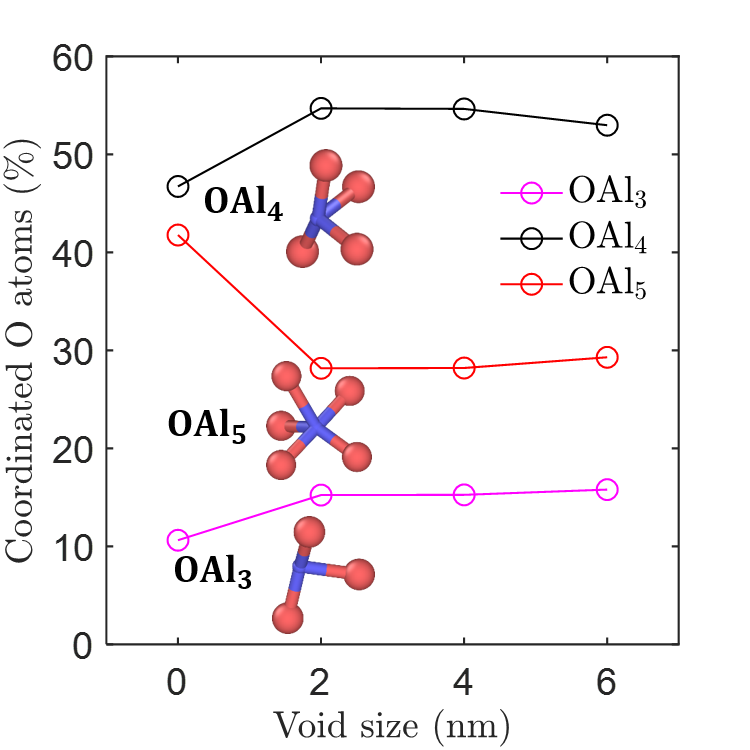}
 		\caption{}
 		\label{fig:Molecular_dynamics_void_size_strain_rate_c}
 	\end{subfigure}	
 	\caption{Effect of nanovoid size on $\mathrm{Al_{2}O_{3}}$ (a) stress-strain behavior with inset figures of the predicted failure pattern for the selected nanovoid size simulation, and coordination number of (b) Al and (c) O atoms at strain of 5\%. Simulations were conducted with a strain rate of $\mathrm{10^{8}~s^{-1}}$ for a simulation cell with a volume of 30 $\mathrm{nm^{3}}$.}
 	\label{fig:Molecular_dynamics_void_size_strain_rate}
 \end{figure}
 The effect of void size on the stress-strain behaviour is depicted in Figure \ref{fig:Molecular_dynamics_void_size_strain_rate_a}.
 In these simulations, the void diameter was varied between 0 and 6 nm.
 Notably, the spall strength for the defect-free $\mathrm{Al_{2}O_{3}}$ is $\sigma_\mathrm{spall}$  = 4.6 GPa. 
 In contrast, the introduction of nanosized voids led to a reduction in spall strength from 4.6 to 4.45 GPa, although the critical strain remained constant at 1.4$\%$.
 Interestingly, a distinct transition is observed when comparing the defect-free case to samples with nanovoids.
 Failure analysis reveals a decrease in the extent of failure as void size increases, as shown in the inset images of Figure \ref{fig:Molecular_dynamics_void_size_strain_rate_a}. 
 This was observed by tracking the void volume, which was visualized using surface meshes.
 Defect-free samples exhibited brittle failure along multiple spall planes due to lack of any weak spots or stress concentration sites.
 Conversely, samples containing nanovoids failed predominantly along a single spall plane, with increasing damage correlated to larger void diameters.
 Additionally, coordination analysis indicated a rise in bond breakage, evident by the decrease in $\mathrm{AlO_{6}}$ and $\mathrm{OAl_{5}}$ and increase in the other coordinated Al and O atoms, with increasing void size for both Al and O atoms as seen in Figures \ref{fig:Molecular_dynamics_void_size_strain_rate_b}---\ref{fig:Molecular_dynamics_void_size_strain_rate_c}.
 Comparatively, nanovoid cases exhibited significantly more bond breakage than the defect-free samples.
 This transition in failure behavior arises directly from the increased void size, a phenomenon well established in the literature. 
 It is known that an increase in initial defect size reduces the stress-carrying capacity of materials, a trend consistent with our findings and previous studies \cite{Chen2022, Chang_Hogan_2023}.
\end{document}